\documentclass[journal]{IEEEtran}

\usepackage{amsmath,amssymb,amsfonts}
\usepackage{booktabs,siunitx,adjustbox}
\usepackage{multirow}
\usepackage[capitalize,noabbrev]{cleveref}
\usepackage[table]{xcolor}
\usepackage{tcolorbox}
\usepackage{xspace}
\usepackage{pifont}
\usepackage{graphicx}
\usepackage{cite}
\usepackage{url}
\usepackage{caption}

\makeatletter
\DeclareRobustCommand\onedot{\futurelet\@let@token\@onedot}
\def\@onedot{\ifx\@let@token.\else.\null\fi\xspace}
\def\eg{\emph{e.g}\onedot} 
\def\ie{\emph{i.e}\onedot}

\makeatother

\newcommand{\best}[1]{\cellcolor{green!25}\textbf{#1}}
\newcommand{\tool}{\textit{CodeSnitch}\xspace}
\newcommand{\citet}[1]{\cite{#1}}

\begin{document}


\title{Are AI Coders Snitches? An Empirical Study of Pretraining Data Detection on Code Large Language Models}

\author{
\IEEEauthorblockN{
Tianlin Li\textsuperscript{1,*},
Yunxiang Wei\textsuperscript{2,*},
Zhiming Li\textsuperscript{3},
Aishan Liu\textsuperscript{1},
Qing Guo\textsuperscript{4},
Qiang Hu\textsuperscript{5},
Xianglong Liu\textsuperscript{1},
Dongning Sun\textsuperscript{3},
Yang Liu\textsuperscript{6},
and Yves Le Traon\textsuperscript{7}
}
\thanks{
    \textsuperscript{1}Beihang University, Beijing, China;
    \textsuperscript{2}Zhejiang University, Hangzhou, China;
    \textsuperscript{3}Peng Cheng Laboratory, Shenzhen, China;
    \textsuperscript{4}Nankai University, Tianjin, China;
    \textsuperscript{5}Tianjin University, Tianjin, China;
    \textsuperscript{6}Nanyang Technological University, Singapore;
    \textsuperscript{7}University of Luxembourg, Luxembourg;
    \textsuperscript{*}Equal Contribution
}
\thanks{
E-mail: tianlin001@buaa.edu.cn; 3230105111@zju.edu.cn; lizhm02@pcl.ac.cn; liuaishan@buaa.edu.cn; tsingqguo@ieee.org; qianghu@tju.edu.cn; xlliu@buaa.edu.cn; sundn@pcl.ac.cn; yangliu@ntu.edu.sg; yves.letraon@uni.lu.
}}

\markboth{}{}

\maketitle

\begin{abstract}
Recent advancements in code large language models (CodeLLMs) have made them essential tools in software engineering.
However, their outputs occasionally include proprietary or sensitive code snippets, indicating non-compliant code usage in training data and thereby raising concerns regarding privacy and intellectual property.
Conducting code pretraining data detection (PDD) is therefore essential for ensuring compliant deployment.
Recent PDD methods have shown promise for natural language data. However, their effectiveness for code data remains largely unclear, especially given code's structured grammar and fundamentally different criteria for evaluating similarity.

To address this gap, in this paper, we systematically evaluate 9 state-of-the-art PDD methods on code data using 8 CodeLLMs.
To enable this evaluation, we construct a comprehensive, function-level source code benchmark \tool, comprising 9,600 code samples across three programming languages, each explicitly labeled to be either included in or excluded from CodeLLM training.
In addition to the evaluation on original \tool, to better assess the effectiveness of existing PDD methods, we propose targeted data mutation strategies to evaluate the robustness of these detection approaches under three different settings.
Based on this study, we propose \textbf{RCM} (Reverse Code Mutation), which mitigates the degradation brought by the evaluated code mutations. This study enhances our understanding of existing PDD methods and lays the groundwork for developing more advanced approaches.
\end{abstract}


\section{Introduction}
\IEEEPARstart{T}{he} advancement of code capabilities has established code large language models (CodeLLMs) as essential productivity tools in software engineering.
Many CodeLLM-based tools have been developed for tasks such as code generation \cite{li2023starcodersourceyou,liu2023codegeneratedchatgptreally} and code repair \cite{10764852,Xia_2023}, achieving successful real-world applications.
For example, Microsoft Copilot, which serves as a coding assistant, has reportedly reached 33 million active users \cite{businessofapps_copilot2025}. CodeLLMs will continue to significantly support software engineering tasks over the long term. \looseness=-1

Despite their impressive performance and promising applications, the use of large-scale pretraining data collected from the web can lead to unintended negative consequences in CodeLLMs.
For example, users frequently report that the code generated by these models may contain proprietary source code or sensitive personal information~\cite{maben2025_privacyleakage,li2024purifying}. Such incidents suggest that large-scale pretraining datasets, often gathered from diverse public sources, may inadequately comply with copyright and privacy regulations, potentially infringing data owners' rights. Moreover, large-scale pretraining datasets have been reported to unintentionally contain evaluation benchmark data \cite{jiang2024investigatingdatacontaminationpretraining}, which significantly compromises the reliability of evaluations.
Facing this situation, \textbf{pretraining data detection} (PDD), which aims to identify whether a given code sample was included in a model's pretraining data, is increasingly important, as successful detection helps protect data owners' rights and reduces the impact of data contamination. 
This problem is a special case of Membership Inference Attacks (MIAs)~\cite{shokri2017membershipinferenceattacksmachine} that focuses on inferring membership in pretraining data rather than in standard task-specific training datasets. 
PDD is particularly challenging because pretraining corpora are typically much larger, and each individual sample is usually processed only once, leaving a much weaker footprint on the learned parameters.

The research community has made significant progress in training data detection for small AI models, formulating it as a binary classification problem \cite{hu2022membership}. Although these approaches are well-developed, they are not applicable to LLMs. 
This limitation primarily stems from their reliance on access to the original training dataset, which is typically unavailable for LLMs due to the proprietary nature of their training data.

Although challenging, it is encouraging to see ongoing efforts in PDD for general LLMs that avoid the need for direct access to pretraining data.
\citet{shi2024detectingpretrainingdatalarge} are the first to study PDD for general LLMs. They propose Min-K\%, which identifies low-probability outlier tokens and averages their log-likelihoods to detect pretraining data based on a set threshold.
\citet{zhang2025minkimprovedbaselinedetecting} extend this method and propose Min-K\%++ by normalizing token log probabilities using their mean and variance, aiming to improve detection sensitivity. Additionally, one recently introduced attack has advanced this field further: RECALL \cite{xie2025recallmembershipinferencerelative} measures the relative change in log-likelihood when conditioning the target text on specific prefixes.

However, despite the commendable progress for general LLMs, it remains unclear whether these methods are effective for CodeLLMs.
One source of the uncertainty arises from the structured nature of code—it adheres to strict grammatical rules while incorporating flexible, human-defined semantics and variable names. 
This can significantly impact the performance of existing PDD methods on code data.
Additionally, the definition of training data membership in code is another factor that may undermine the effectiveness of existing PDD methods. 
Existing PDD methods often rely on a narrow definition—treating only verbatim equivalence (\ie, exact matches) with original training data as membership, while ignoring semantically equivalent variations. However, code data requires a more nuanced training data membership definition. 
This is because, in software engineering, well-established code clone standards (Type I–Type IV) \cite{roy2009comparison,10967509} are used to characterize related code fragments, which is directly relevant to the definition of membership.
These standards imply broader interpretations of membership beyond exact matches. 

\textbf{Our work.} We conduct an in-depth and extensive empirical study to evaluate how existing PDD methods, originally validated on general LLMs, perform on code data.
Specifically, we first curate a function-level source code dataset that includes both data commonly used to train CodeLLMs and data excluded from their training.
We further design data mutation strategies based on Type-1 to Type-3 code-clone definitions and their hybrid combination, and evaluate whether existing methods can maintain reliable performance under these mutations.
In this study, we identify the limitations of existing approaches.
Moreover, we propose RCM (Reverse Code Mutation), which mitigates the degradation brought by the evaluated code mutations.

\textbf{Contributions.} In summary, the research makes the following contributions:
\ding{182} We develop \tool, a function-level code dataset spanning three programming languages, which includes 9,600 samples comprising both training and non-training data, providing a valuable resource for future research.
\ding{183}  We conduct an in-depth study of nine state-of-the-art PDD methods and eight CodeLLMs to evaluate their performance on code data.
\ding{184} We propose RCM to  mitigate the degradation brought by the evaluated code mutations.
\ding{185}  We have released \tool and the evaluation pipeline in \cite{website}. We have advanced the understanding of PDD, establishing a foundation for the compliant use of code data.

\section{Background}
In this section, we provide a brief introduction to the relevant background, including CodeLLM training, the training data detection task, and code clone definitions.

\subsection{CodeLLM Training}
CodeLLMs are trained on a diverse collection of languages, including both natural and programming languages \cite{li2023starcodersourceyou}. They have demonstrated outstanding performance in various code generation tasks, such as vulnerability patch generation and code completion \cite{yang2024acecodereinforcementlearningframework}. Moreover, CodeLLMs show great potential in addressing complex software engineering challenges, with established benchmarks playing as a vital role in evaluating their capabilities \cite{wang2025softwaredevelopmentlifecycle}.

Large datasets used to train CodeLLMs often contain copyrighted code, which can lead to models memorizing and reproducing segments of this code without permission \cite{yu2023codeipprompt,li2024purifying,wang2024trustworthyllmscodedatacentric}. This raises concerns about whether the models are being trained on copyrighted material without the permission of the original authors.
Such large datasets may also include data present in the evaluation benchmark, thereby undermining the fairness of the evaluation.

\subsection{Training Data Detection}
\label{bg_PDD}

Training data detection refers to methods that determine whether a specific data record was used in the training of an AI model, a concept first introduced by \citet{shokri2017membershipinferenceattacksmachine}.
While these methods may not directly result in privacy breaches, they can help identify whether a model was trained using data that violates data usage regulations.
Specifically, researchers formulate this task as a binary classification problem \cite{hu2022membership}.


Formally, given a sequence of tokens in a data instance \( x = x_1, x_2, \ldots, x_N \), and a pre-trained auto-regressive LLM $M$ trained on a dataset $\mathcal{D}_t$,
the goal is to infer whether $x \in \mathcal{D}_t$ (\ie, whether $x$ is part of the training data or not). These detection methods typically leverage a scoring function $s(x; M)$ that computes a score for each input. A threshold is then applied to the score to yield a binary prediction:
\[
\text{prediction}(x, M) =
\begin{cases}
1 & (x \in \mathcal{D}_t),\quad s(x; M) \geq \lambda \\
0 & (x \notin \mathcal{D}_t),\quad s(x; M) < \lambda
\end{cases}
,
\]
where $\lambda$ is a case-dependent threshold.
Note that, as defined above, existing definitions of training data membership rely on a \emph{verbatim assumption}—that is, a sample is considered part of the training set only if it appears exactly as it did during training.

Here, we briefly introduce seven widely used PDD methods for LLMs, followed by two recent code-specific methods.

\ding{182} \textbf{Perplexity (PPL) \cite{yeom2018privacyriskmachinelearning}}: Utilizes the perplexity of a text, as computed by the LLM, to determine its membership. A lower perplexity suggests the text is a member. The score function is defined as:
    $$s(x; M) = \log p(x; M),$$
    where $\log p(x; M)$ = $\frac 1 N \sum_1^N\log p(x_i|x_1,\ldots,x_{i-1}; M)$ is the average log generation probability of data instance $x$ and $N$ is the length. Consider the example $x=\texttt{def~add(a, b): return a + b}$, PPL scores the tokens of $x$ directly and averages their conditional log probabilities.


\ding{183} \textbf{zlib  \cite{274574} }: Employs zlib compression to estimate the entropy of a text, which is then used to calibrate the perplexity score and mitigate the influence of text complexity. The score function is defined as:
    $$s(x; M) = \frac{\log p(x; M)}{\text{length}(\texttt{zlib.compress}(x))},$$
    where $\text{length()}$ is the length function and $\texttt{zlib.compress()}$ is the zlib compression function. Consider the example $x=\texttt{def~add(a, b): return a + b}$, zlib divides its model score by the length of its compressed representation to calibrate for code complexity.

\ding{184} \textbf{Min-K\% \cite{shi2024detectingpretrainingdatalarge}}: This method calculates the metric based only on the k\% of tokens with the lowest generation probabilities, hypothesizing that unmemorized sequences are characterized by a few tokens with very low confidence. This method selects the \( k\% \) of tokens from \( x \) with the minimum token probability to form a set, Min-K\%(\( x \)), and computes the average log-likelihood of the tokens in this set:
    \[
s(x; M) = \frac{1}{E} \sum_{x_i \in \text{Min-K\%}(x)} \log p(x_i \mid x_1, \ldots, x_{i-1}; M),
\]
where \( E \) typically represents the number of tokens in the set \(\text{Min-K\%}(x)\). Consider the example $x=\texttt{def~add(a, b): return a + b}$, Min-K\% uses only the selected low-probability tokens such as $\texttt{add}$, while Min-K\%++ normalizes those same token-level scores.

\ding{185} \textbf{Min-K\%++ \cite{zhang2025minkimprovedbaselinedetecting}}:
An extension of Min-K\% that normalizes token log probabilities using their mean and variance. This reduces bias from inherently high- or low-probability tokens, providing a more robust measure of low-confidence or anomalous tokens for detection tasks.

\ding{186} \textbf{Reference (Ref)  \cite{274574} }: Calibrates the loss using a smaller, reference model. The core assumption is that member data is significantly better memorized by the larger target model, a phenomenon not observed for non-member data. The score function is defined as:
    $$s(x; M, {R}) = \log p(x; M) - \log p(x; {R}),$$
    where ${R}$ is a selected reference model. For example, $M=\texttt{StarCoder2-7B}$ is the larger target model and $R=\texttt{StarCoder2-3B}$ is the smaller reference model from the same family. Ref evaluates $x$ with both models and subtracts the score from $R$ from the score from $M$.

\ding{187} \textbf{Neighbor \cite{mattern2023membershipinferenceattackslanguage}}: This method first creates neighborhood samples by making slight modifications to the original input. It assumes that the loss of a member sample will be substantially lower than that of its neighbors, whereas this disparity will be less pronounced for non-member samples. The score function is defined as:
    \begin{align*}
    s(x; M) &= \log p(x; M) -
    \frac{1}{k} \sum_{i=1}^{k} \log p(\text{neighbor}_i; M),
    \end{align*}
    where $\text{neighbor}_i$ is a modified sample based on $x$. For example, for $x=\texttt{def~add(a, b): return a + b}$, an infilling model can receive $\texttt{def~add(a, b): return <mask>}$ and fill the mask with $\texttt{a - b}$, producing the neighbor $\texttt{def~add(a, b): return a - b}$. The target model then compares the score of $x$ with the average score of these generated neighbors.

\ding{188} \textbf{ReCaLL} \cite{xie2025recallmembershipinferencerelative}: This method computes the change in loss for a sample when prefixed with a non-member context. The hypothesis is that the loss of member data will decrease more significantly with the prefix compared to non-member data. The score function is defined as:
\begin{align*}
s(x; M) &= \log p(x; M) - \log p(x \mid \text{non-member prefix}; M).
\end{align*}
For example, let $x=\texttt{def~add(a, b): return a + b}$ and let $p=\texttt{def~subtract(a, b): return a - b}$ be a held-out non-member prefix. ReCaLL compares the loss of $x$ alone with the loss of $x$ after concatenating $p$ and $x$, their difference is the ReCaLL score.

{\ding{189} \textbf{CodeMI \cite{wan2026codemi}}: Trains shadow code-completion models to obtain token-rank features and uses these features to train an attack classifier that predicts whether a candidate sample was used to train the target model.

\ding{190} \textbf{AdvPrompt-MIA \cite{jiang2025advprompt}}: Applies semantics-preserving transformations to code prompts and trains an attack classifier using the resulting changes in completion similarity and perplexity.}
Note that \cite{yang2024gotchamodelusescode} proposes an insightful method for evaluating membership leakage risks in code models like CodeBERT \cite{feng2020codebertpretrainedmodelprogramming}. However, their approach also relies on training a surrogate model, which is not applicable to our study's focus on CodeLLMs, whose training data is typically proprietary.
\citet{yang2025semantically} investigates membership inference risks, focusing on the data used to fine-tune CodeLLMs, which is fundamentally different from our motivation and research focus.
Thus, in this paper, we thoroughly study the effectiveness of the aforementioned methods when applied to code data.

Based on how their membership scores are constructed, we group the seven general-purpose PDD methods into \emph{direct-score} and \emph{comparison-based} methods. Direct-score methods (PPL, zlib, Min-K\%, and Min-K\%++) aggregate or normalize the target model's token likelihoods for the candidate code. Comparison-based methods construct relative scores by comparing the candidate's score with its score under a reference model (Ref), the scores of generated neighbors (Neighbor), or its score with a non-member prefix context (ReCaLL).

\subsection{Code Clone Types}

Code cloning \cite{roy2009comparison,10967509} refers to the presence of identical or similar source code fragments within a codebase. The definition of code clones provides a clear and structured analytical framework that can assist in the assessment and determination of non-compliant code usage.
Such clones are typically classified based on their degree of similarity, leading to the following four types:
\ding{182} \textbf{Type-1 (Textual similarity):} Code fragments are identical except for differences in whitespace, layout, and comments.
\ding{183} \textbf{Type-2 (Lexical similarity):} Code fragments are identical except for variations in identifier names, literals, and lexical values, in addition to the differences associated with Type-1 clones.
\ding{184} \textbf{Type-3 (Syntactic similarity):} Code fragments are syntactically similar but differ at the statement level. Beyond variations observed in Type-1 and Type-2 clones, these fragments may contain added, modified, or removed statements.
Researchers typically use a similarity score as a threshold, defined as the ratio of unchanged lines to the total number of lines.
\ding{185} \textbf{Type-4 (Semantic similarity):} Code fragments are syntactically dissimilar but achieve identical functionality.

\begin{figure}[htb]
    \centering
    \includegraphics[width=\linewidth]{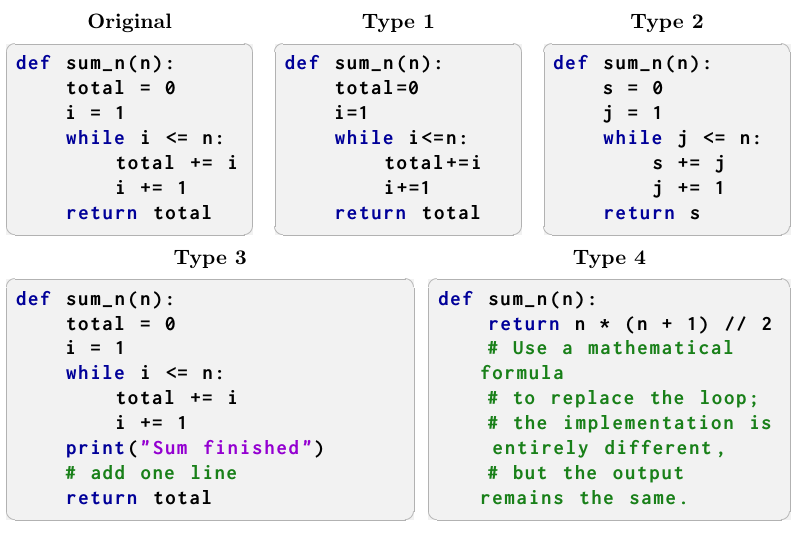}
    \caption{The Four Types of Code Clones}
    \label{pic:codeclone}
\end{figure}
To visually illustrate these four types of code clones, Fig. \ref{pic:codeclone} provides example scenarios derived from an original code fragment that implements a cumulative summation function.





In this paper, we use the Type-1 to Type-3 definitions to systematically investigate how existing PDD methods perform when applied to CodeLLMs.

\section{Motivation}
Here, we first analyze the necessity of conducting this study—that is, why existing PDD methods designed for natural language (NL) data may not perform as effectively on code data.



\subsection{Perspective 1: Distinct Code Structure}
The structured nature of code raises questions about the effectiveness of existing methods designed for NL. Specifically, code adheres to strict grammatical rules while incorporating flexible, human-defined semantics and variable names.

Tokens related to grammar and variable names serve distinctly different roles in determining whether code data is part of the training dataset. For example, in Python code, the token ``in'' is likely to occur after the token ``for'' due to grammatical structure. Theoretically, the ``in'' token is less significant for membership detection since the code follows a consistent grammar that dictates ``in'' will occur after ``for''. Therefore, methods like perplexity (PPL), which treat all tokens in a sample equally, may not be effective for code.
\looseness=-1

\subsection{Perspective 2: The Definition of Training Data Membership}
\textbf{The existing training data membership definition has inherent limitations.} In existing work on detecting training data membership for NL, membership is typically defined under a verbatim-based assumption, \ie, a sample is considered part of the training set only if it appears exactly as it did during training.
For example, if only the sentence ``The CEO of Apple is Tim Cook.'' appears in the training data, semantically equivalent variants—such as ``The current CEO of Apple is Tim Cook.''—would not be recognized as part of the training set. 
However, this assumption is limited in scope. A data collector could easily modify a sentence to evade PDD techniques based on this assumption. As a result, methods relying on this assumption offer little practical value for downstream applications such as copyright infringement identification. Nevertheless, given the inherent difficulty of precisely defining training data membership for NL data, existing work generally continues to adhere to this assumption.

\textbf{This limitation also applies to code training data.} This verbatim-based assumption is also limited when determining training data membership in the context of source code. Consider, for example, the scenario of applying PDD methods for copyright protection purposes. Suppose the original code depicted in Fig. \ref{pic:codeclone} is copyrighted material. If a detection method recognizes training data membership only by exact verbatim matches and overlooks Type-1 code clones—such as versions modified merely through whitespace or formatting changes—it becomes easy for data collectors to evade detection by simply reformatting the original code, thereby circumventing copyright restrictions.

\textbf{Our idea to address this limitation for code training data.} 
Unlike NL, which relies on the verbatim definition, the code domain has clearly defined clone types that can be leveraged for a more precise analysis and detection of copyright infringement. Therefore, we propose utilizing these well-established code clone types to redefine the notion of training data membership for code.

\begin{figure*}[t]
    \centering
    \includegraphics[width=\linewidth]{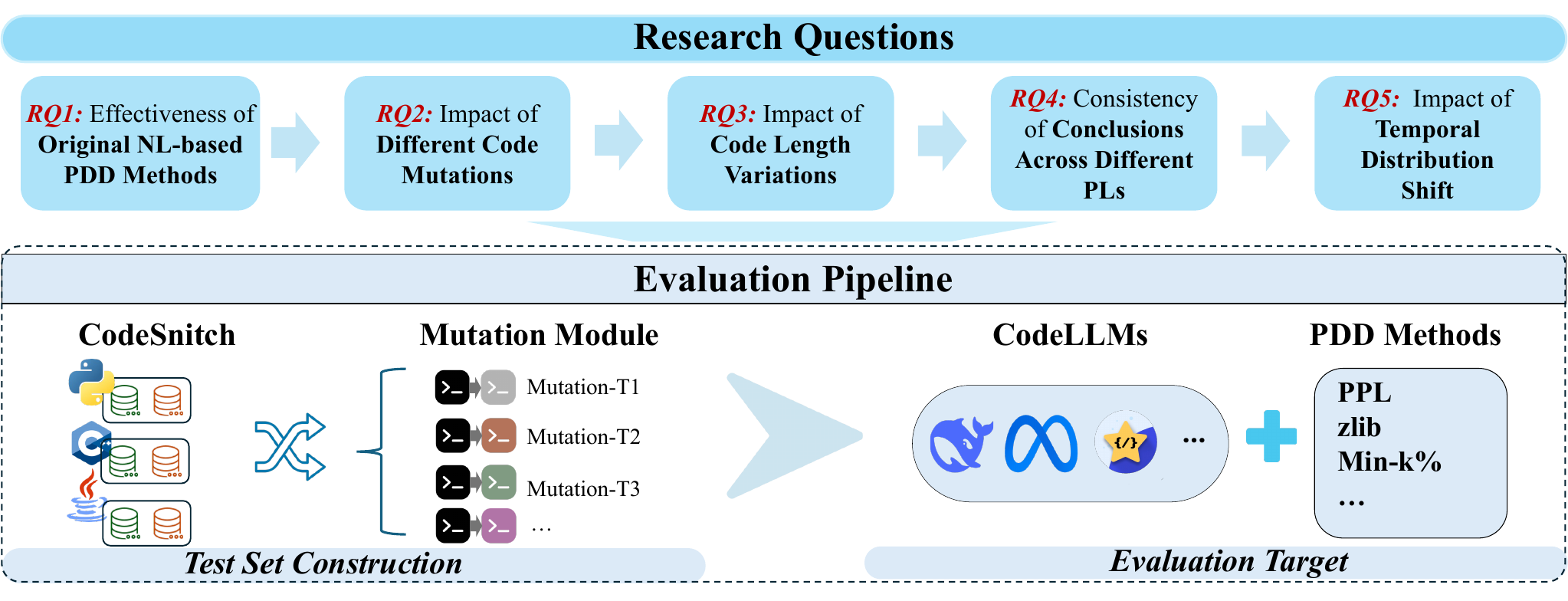}
    \caption{Overview of Our Study} 
    \label{pic:overview} 
\end{figure*}

\section{Study Design}

\subsection{Research Questions}
\label{subsec:RQ}
The major focus of this paper is to systematically understand the effectiveness of the 
state-of-the-art PDD approaches within the context of CodeLLMs. To achieve a comprehensive and systematic understanding, we conduct evaluations in terms of five \emph{research questions} (RQ). The overview of our proposed RQs are shown in Fig. \ref{pic:overview}. In the following content, we will introduce the detailed research questions adopted and the corresponding experimental design for their quantitative evaluations.

\textit{\textbf{RQ1: Effectiveness of Existing PDD Methods.}} For the first research question (RQ1), we aim to comprehensively evaluate the effectiveness of state-of-the-art PDD methods. The evaluated methods include both approaches originally designed for natural language and recent code-specific approaches. To enable this evaluation, we construct a comprehensive, function-level source code benchmark \tool, covering three programming languages (PL): Python, Java, and C++.
We adopt eight representative CodeLLMs as backbones and nine PDD methods for evaluation\footnote{Note that our implementation pipeline is highly reusable and extensible, and is readily available for the future incorporation of other Transformer-based foundation models.}.

\textit{{\textbf{RQ2: Impact of Different Code Mutations.}}}
For RQ2, we aim to thoroughly investigate the impact of different code mutations of the state-of-the-art PDD techniques. Specifically, unlike NL, where it is generally assumed that only verbatim repetition leads to non-compliance, source code commonly exhibits various forms of code clones (\ie, code fragments that are related by textual, lexical, syntactic, or semantic similarity). To account for this, we have implemented a mutation module that augments the benchmark by systematically applying four types of mutation strategies, each corresponding to one of the standard clone types (\ie~Type 1-3) and their hybrid combination. We illustrate a fine-grained analysis of the performance of different PDD methods under the evaluated code mutations.

In order to comprehensively evaluate the effectiveness and robustness of the PDD methods, we further develop a mutation module that introduces several mutation strategies that correspond to different types of code clones (denoted as Mutation-T$l$, $l\in[1,3]$). The details of the proposed mutation strategies are as follows:
\ding{182} \textbf{Mutation-T1 (Formatting):} We introduce stylistic mutations by reformatting code snippets using a formatter with randomly generated style parameters. This process alters whitespace characters, such as indentation and spacing, without affecting the code's logic.
\ding{183} \textbf{Mutation-T2 (Lexical):} We define two sub-strategies: 
    \textbf{Mutation-T2a} involves renaming identifiers with randomly selected synonyms. \textbf{Mutation-T2b}, designed to simulate a worst-case scenario, replaces all identifiers with random 8-character strings, removing semantic meaning from them.
\ding{184} \textbf{Mutation-T3 (Syntactic):} We apply statement-level transformations using predefined, semantics-preserving patterns to replace statements with their functional equivalents. Following prior work, we quantify the degree of mutation for Type-3 clones using a similarity score, defined as the ratio of unchanged lines to the total number of lines. We experiment with similarity levels of 0.9, 0.7, and 0.5.
\ding{185} \textbf{Hybrid}: To simulate a more complex transformation, we also create a hybrid mutation by applying a Mutation-T3 mutation (0.8 similarity) on top of a Mutation-T2a (identifier renaming) mutation. We use 0.8 as a fixed intermediate Type-3 similarity level so that the hybrid applies a controlled statement-level change in addition to lexical renaming.
 


\textit{{\textbf{RQ3: Impact of the Code Length Variations.}}}
In practice, source code can vary significantly in length, from small utility functions to large, multi-line implementations. For RQ3, we aim to investigate whether and how the length of source code snippets influences the detection performance of the existing PDD methods. To enable this analysis, we partition our benchmark into five subsets based on code length, creating distinct test sets for functions with token counts in the ranges of 100-200, 200-300, 300-400, 400-500, and 500-600. We then systematically examine how detection results change across these subsets to determine the robustness of each method with respect to varying code lengths.

\textit{{\textbf{RQ4: Consistency of
Conclusions Across
Different PLs.}}}
For RQ4, we assess whether the conclusions drawn from the preceding evaluations are consistent across different programming languages. Since programming languages vary in syntax, idiomatic usage, and structural conventions, this may influence the behavior of PDD techniques. To this end, we compare the results obtained for Python, Java, and C++ separately on all the provided types of code mutation. This helps us understand to what extent the effectiveness and limitations of the evaluated methods generalize across programming languages within the CodeLLM domain.

\textit{{\textbf{RQ5: Sensitivity to Temporal Distribution Shift.}}}
For RQ5, we investigate whether the performance of PDD methods changes as the temporal distance between member and non-member data increases. Prior work has highlighted the influence of temporal differences between member and non-member data on membership-inference evaluations~\cite{duan2024do}. The original member data provide high-confidence member labels based on existing pretraining corpora. We therefore retain these data as the member source and collect additional non-member data from two later temporal windows.

\subsection{Evaluation Pipeline Setup}
\subsubsection{Construction of \tool}
\tool ($\mathcal{D}$) consists of a member dataset ($\mathcal{D}_m$) and a non-member dataset ($\mathcal{D}_{nm}$), \ie, $\mathcal{D} = \{\mathcal{D}_{m}, \mathcal{D}_{nm}\}$. To construct the member dataset $\mathcal{D}_m$, we directly sample function-level code from the publicly accessible \texttt{the-stack} corpus \cite{kocetkov2022stack3tbpermissively} and the \texttt{deepseek-coder} corpus \cite{guo2024deepseekcoderlargelanguagemodel}. The Stack is a documented pretraining source for the StarCoder family, while the DeepSeek-Coder corpus is the documented pretraining source for the DeepSeekCoder family. All selected samples predate the release of the evaluated models. Thus, the corpus provenance provides high-confidence membership evidence for the corresponding model families, but this evidence does not automatically transfer to other CodeLLM families. In particular, the public training documentation does not establish that the CodeLlama models were trained on either source.

For the non-member dataset $\mathcal{D}_{nm}$, we collect data from highly starred GitHub repositories (more than 10{,}000 stars) created after January 1, 2024, as all models in our study were released before this date. Representative examples include \texttt{browser-use}~\cite{browser_use}, \texttt{OpenHands}~\cite{openhands}, and \texttt{MinerU}~\cite{mineru}. We maintain a list of APIs that first appear in these post-cutoff commits and extract functions that explicitly call a listed API; we also extract functions from recent commits made after 2024. Representative API examples include \texttt{marimo}~\cite{marimo}.
The rationale is that code using new APIs is unlikely to appear in the training data of pre-released models. Moreover, code from recent commits—often addressing new requirements—is less likely to consist of boilerplate templates that models may have learned.

For each programming language, we construct 1{,}600 function-level samples for both $\mathcal{D}_{m}$ and $\mathcal{D}_{nm}$ to keep a balanced evaluation setting, resulting in 9{,}600 samples across the three languages. During sample extraction, we use AST-based parsing to retain context-independent functions (\eg, functions whose bodies do not rely on external classes), thereby reducing potential syntax dependencies on surrounding code. We further validate that all extracted functions are syntactically correct and discard any samples with parsing errors to ensure snippet quality. Finally, we perform function-level deduplication using function signatures within and across $\mathcal{D}_{m}$ and $\mathcal{D}_{nm}$ to remove exact overlaps. To validate the quality of the deduplication, we conduct a near-duplicate contamination audit using normalized representations. No cross-set duplicates are found under these checks, corresponding to a 0.0\% contamination rate and confirming that the constructed sets are free of detectable duplicate contamination.

Note that we set the fixed cutoff date as January 1, 2024, for the scope of our study. However, the construction method supports flexible cutoff dates, enabling the study of more advanced models.


\subsubsection{Backbone CodeLLMs}
We select eight representative CodeLLMs from three model families, spanning a range of model sizes, as our backbone models for evaluation: the StarCoder family \cite{allal2023santacoderdontreachstars, li2023starcodersourceyou}, including \texttt{SantaCoder} (1.1B), \texttt{StarCoder2-3B}, and \texttt{StarCoder2-7B}; the CodeLlama family \cite{rozière2024codellamaopenfoundation}, including \texttt{CodeLlama-7B} and \texttt{CodeLlama-13B}; and the DeepSeekCoder family \cite{guo2024deepseekcoderlargelanguagemodel}, including \texttt{DeepSeekCoder-1.3B},
\texttt{DeepSeekCoder-6.7B}, and \texttt{DeepSeekCoder-33B}, ensuring a comprehensive and practical assessment.

\subsubsection{Evaluated Methods}
In this paper, we investigate seven widely used methods—PPL, zlib, Min-K\%, Min-K\%++, Ref, Neighbor, and ReCaLL—and two recent code-specific methods, CodeMI and AdvPrompt-MIA. Please refer to \Cref{bg_PDD} for the detailed introduction.

\subsubsection{Evaluation Metrics}
Following previous PDD work, we evaluate the performance of the detection methods using the Receiver Operating Characteristic (ROC) curve \cite{metz1978basic}, reporting the Area Under the Curve (AUC) score.
The AUC metric is widely used to assess binary classification tasks, such as the detection of training data \cite{zhang2024min}. 
To quantify uncertainty in AUC differences, we use a paired, stratified nonparametric bootstrap over member and non-member functions and report two-sided 95\% confidence intervals for $\Delta\mathrm{AUC}$.

\subsubsection{Implementation Details}
In this study, we treat code as textual input and apply PDD methods to CodeLLMs to evaluate their performance.
For \textbf{Min-K\%} and \textbf{Min-K\%++}, following prior work, we set $K=20$ and $K=30$, respectively~\cite{shi2024detectingpretrainingdatalarge,zhang2025minkimprovedbaselinedetecting}.
For \textbf{ReCaLL}, following the original paper's suggestion, we randomly sample 12 non-member code snippets from $\mathcal{D}_{nm}$, which are then excluded from the test set, to serve as prefixes. Then we compute the loss of each sample with the non-member context and without it to get the difference.
We put more details in \cite{website}.
For the \textbf{Neighbor} method, we adhere to the original methodology, using the code-oriented \texttt{CodeT5+} 220M model~\cite{wang-etal-2023-codet5} with random masking to generate neighborhood samples.
For the \textbf{Ref} method, we use \texttt{StarCoder2-7B}, \texttt{DeepSeekCoder-6.7B,33B}, and \texttt{CodeLlama-13B} as the primary models, with \texttt{StarCoder2-3B}, \texttt{DeepSeekCoder-1.3B}, and \texttt{CodeLlama-7B} serving as their respective reference models.
{For \textbf{CodeMI}, following its original protocol, we train two CodeGPT shadow models on an auxiliary Python corpus and use token-rank histograms from their completions to train an attack classifier. For \textbf{AdvPrompt-MIA}, we generate completions from the original function prefix and 11 semantics-preserving variants, then train a target-specific classifier using completion-similarity and perplexity features. Samples used to train either attack classifier are excluded from its evaluation.}
To ensure the stability and reliability of our results, we conduct ten independent experiments with different random seeds and report the average AUC scores.

\begin{table*}[ht]
\centering
\small
\setlength{\tabcolsep}{8pt}
\renewcommand{\arraystretch}{1.1}
\caption{AUC comparison across models and methods. Avg is the average AUC score across all models. The “–” indicates that no result is available due to the absence of a reference model for the smaller models.}
\begin{adjustbox}{width=\textwidth}
\begin{tabular}{lccccccccccc}
\toprule
Model & PPL & Zlib & Min-20\% & Min-30\% & Min-20\%++ & Min-30\%++ & Ref & Neighbor & CodeMI & AdvPrompt & ReCaLL \\
\midrule
\midrule
SantaCoder 1.1B & 0.594 & 0.573 & 0.583 & 0.589 & 0.553 & 0.583 & \text{-} & 0.573 & 0.503 & 0.581 & \textbf{0.790} \\
StarCoder2 3B & 0.596 & 0.573 & 0.583 & 0.591 & 0.552 & 0.589 & \text{-} & 0.563 & 0.504 & 0.589 & \textbf{0.796} \\
StarCoder2 7B & 0.608 & 0.588 & 0.597 & 0.604 & 0.562 & 0.601 & 0.597 & 0.571 & 0.501 & 0.587 & \textbf{0.830} \\
CodeLlama 7B & 0.650 & 0.628 & 0.646 & 0.646 & 0.589 & 0.610 & \text{-} & 0.592 & 0.533 & 0.620 & \textbf{0.801} \\
CodeLlama 13B & 0.653 & 0.631 & 0.645 & 0.648 & 0.592 & 0.619 & 0.494 & 0.594 & 0.531 & 0.516 & \textbf{0.813} \\
DeepSeekCoder 1.3B & 0.568 & 0.542 & 0.568 & 0.570 & 0.544 & 0.564 & \text{-} & 0.543 & 0.499 & 0.526 & \textbf{0.741} \\
DeepSeekCoder 6.7B & 0.593 & 0.568 & 0.588 & 0.590 & 0.563 & 0.578 & 0.585 & 0.565 & 0.502 & 0.496 & \textbf{0.793} \\
DeepSeekCoder 33B & 0.634 & 0.611 & 0.625 & 0.629 & 0.595 & 0.618 & 0.656 & 0.588 & 0.513 & 0.604 & \textbf{0.806} \\
\midrule
Avg & 0.612 & 0.589 & 0.604 & 0.608 & 0.569 & 0.595 & 0.583 & 0.574 & 0.511 & 0.565 & 0.796 \\
\bottomrule
\end{tabular}
\end{adjustbox}

\label{tab:all_model_auc}
\end{table*}

\section{Evaluation Results}


\subsection{RQ1: Effectiveness on \tool}

To answer RQ1, we evaluate nine PDD methods on \tool $\mathcal{D}$ (Python). The results, measured by AUC scores, are presented in Table~\ref{tab:all_model_auc}. To complement AUC, we also compute the true positive rate at a 5\% false positive rate (TPR@5\%FPR) from the same ROC curves.

\textbf{Comparison among different methods.}
As presented in Table~\ref{tab:all_model_auc}, we observe that most existing methods demonstrate limited effectiveness, with scores below or around 0.6 in 8 out of 9 methods. For instance, Min-K\%++, which is effective on natural language, achieves an AUC of merely 0.595. The two code-specific methods also remain limited: CodeMI and AdvPrompt-MIA achieve average AUC scores of 0.511 and 0.565, respectively. This indicates that the distinct structural properties of code, as opposed to general text, pose unique challenges for established detection methods.

The zlib method, which normalizes perplexity by the sample's compressed size to approximate its entropy, underperforms the PPL baseline, with its AUC dropping from 0.612 to 0.589. This suggests that calibrating perplexity using compression entropy, a viable approach for general text, is ineffective for code, as the structural complexity of code is not well-captured by its compressibility.

Similarly, the Neighbor method contrasts the loss of an original input with that of its perturbed neighbors. With code-oriented CodeT5+ infilling, Neighbor obtains an average AUC of 0.574, which is above random guessing but remains below PPL and ReCaLL. Even with a code-oriented generator, the Neighbor score provides only a limited signal for distinguishing member and non-member code.

The Min-K\% and Min-K\%++ methods, which focus the detection signal on the K\% of tokens with the lowest generation probabilities, also don't outperform PPL, with scores of 0.608 and 0.595 (for K=30), respectively. We conduct a detailed analysis and identify a potential cause: tokens corresponding to the first-time declaration of variable names consistently exhibit low generation probabilities in \textit{both} member and non-member snippets. Since Min-K\% methods preferentially select these low-probability tokens, their ability to discriminate between the two sets is diminished, likely explaining their limited effectiveness on code.

The reference-based method, Ref, which calibrates a target model's loss against that of a smaller reference model, also fails to demonstrate strong performance. We hypothesize this is due to the strong generalization capabilities of code models on common logical and syntactic structures. As a result, both member and non-member snippets exhibit a similar drop in loss when evaluated on a larger model compared to a smaller reference model, thereby reducing the effectiveness of the detection method. Because Ref requires an available reference model pair, its reported Avg is computed over the four models for which such a pair is available and should be interpreted within this subset.

Notably, the ReCaLL method stands out, achieving a superior average AUC score of 0.796. This method calculates the change in loss when a sample is prepended with a few non-member prefixes. This superior performance is likely because the 12 non-member prefixes selected for ReCaLL contain time-sensitive information (post-dating 2024). For non-member data, drawn from a similar temporal distribution, this contextual prefix effectively reduces the snippet's perplexity. Conversely, for member data, this non-member prefix disrupts the memorized sequence, increasing perplexity. The resulting ratio of loss with the prefix to the original loss thus serves as a powerful discriminator. Across the eight evaluated CodeLLMs, ReCaLL exceeds PPL by $\Delta\mathrm{AUC}=0.184$ (95\% CI $[0.178, 0.189]$), and the interval excludes zero.

For completeness, we also report the mean TPR at 5\% FPR across the RQ1 evaluation runs. Among the direct-score methods, PPL, zlib, Min-20\%, Min-30\%, Min-20\%++, and Min-30\%++ achieve mean TPR@5\%FPR values of 0.106, 0.095, 0.099, 0.103, 0.116, and 0.115, respectively. For the comparison-based methods, Ref achieves 0.140 on its four available pairs, Neighbor achieves 0.090, and ReCaLL achieves 0.281. The code-specific methods obtain 0.000 for CodeMI and 0.090 for AdvPrompt-MIA. Thus, the low-FPR metric shows the same overall pattern as AUC: ReCaLL provides the strongest detection signal among the evaluated methods.

\textbf{Comparison among different models.}
For the model families we evaluated—\texttt{CodeLlama}, \texttt{DeepSeekCoder}, and \texttt{StarCoder2}—the performance of most PDD methods improves as model size increases. For instance, with the PPL baseline on the \texttt{DeepSeekCoder} series, AUC scores consistently rise from 0.568 to 0.593 and to 0.634 with increasing model scale. This scaling trend holds for other methods and model families.
This may be attributed to the fact that larger models tend to memorize training data more strongly \cite{liu2024probinglanguagemodelspretraining,shi2024detectingpretrainingdatalarge}, thereby improving detection performance.
Furthermore, the relative performance ranking of the methods remains consistent across all tested models (\eg, PPL > Min\% > zlib), suggesting the fundamental efficacy of these detection strategies is largely model-agnostic.

\begin{tcolorbox}[size=title]
	{\textbf{Answer to RQ1: ReCaLL consistently achieves the best performance across all models, with a superior average AUC score of 0.796, whereas the average performance of other methods is only around 0.6. Larger models have higher detection accuracy. The results highlight that more advanced PDD methods are needed for code data.}}
\end{tcolorbox}

\begin{table*}[htbp]
  \centering
  \setlength{\tabcolsep}{4pt}
  \renewcommand{\arraystretch}{1.1}
    \caption{AUC scores under Setting 1 (Member-only Mutation). The ``–'' indicates that no result is available due to the absence of a reference model for the 1.1B SantaCoder. $\text{MT1}$, $\text{MT2}$, $\text{MT3}$, and $\text{Hyb}$ are abbreviations for Mutation-T1, Mutation-T2, Mutation-T3, and Hybrid, respectively. $\text{MT3}_{0.9}$ is an abbreviation for Mutation-T3 with a similarity threshold of 0.9.}
  \begin{minipage}[t]{0.47\textwidth}
    \centering
    \textbf{SantaCoder 1.1B}\par
    \vspace{2pt}
    \begin{adjustbox}{width=\linewidth}
    \begin{tabular}{l*{7}{S}}
      \toprule
      \multirow{1}{*}{Method} 
       & {Org} & {MT1} & {MT2a} & $\text{MT3}_{0.9}$ & $\text{MT3}_{0.7}$ & $\text{MT3}_{0.5}$ & {Hyb} \\
      \midrule
      PPL        & 0.594 & 0.586 & 0.545 & 0.508 & 0.485 & 0.471 & 0.461 \\
      Zlib        & 0.573 & 0.569 & 0.522 & 0.494 & 0.475 & 0.466 & 0.450 \\
      Min-20\%        & 0.583 & 0.572 & 0.526 & 0.493 & 0.479 & 0.464 & 0.452 \\
      Min-30\%        & 0.589 & 0.579 & 0.535 & 0.499 & 0.481 & 0.465 & 0.454 \\
      Min-20\%++        & 0.553 & 0.558 & 0.510 & 0.462 & 0.439 & 0.425 & 0.408 \\
      Min-30\%++        & 0.583 & 0.585 & 0.549 & 0.497 & 0.463 & 0.452 & 0.437 \\
      Ref        & \text{-} & \text{-} & \text{-} & \text{-} & \text{-} & \text{-} & \text{-} \\
      Neighbor        & 0.573 & 0.563 & 0.551 & 0.526 & 0.514 & 0.499 & 0.498 \\
      CodeMI & 0.503 & 0.490 & 0.476 & 0.454 & 0.455 & 0.455 & 0.462 \\
      AdvPrompt & 0.581 & 0.608 & 0.616 & 0.617 & 0.609 & 0.603 & 0.610 \\
      ReCaLL        & \best{0.790} & \best{0.787} & \best{0.751} & \best{0.714} & \best{0.679} & \best{0.667} & \best{0.640} \\
      \bottomrule
    \end{tabular}
    \end{adjustbox}
  \end{minipage}
  \hfill
  \begin{minipage}[t]{0.47\textwidth}
    \centering
    \textbf{StarCoder2 7B}\par
    \vspace{2pt}
    \begin{adjustbox}{width=\linewidth}
    \begin{tabular}{l*{7}{S}}
      \toprule
     \multirow{1}{*}{Method} 
       & {Org} & {MT1} & {MT2a} & $\text{MT3}_{0.9}$ & $\text{MT3}_{0.7}$ & $\text{MT3}_{0.5}$ & {Hyb}  \\
      \midrule
      PPL        & 0.608 & 0.600 & 0.548 & 0.517 & 0.494 & 0.483 & 0.466 \\
      Zlib        & 0.588 & 0.586 & 0.525 & 0.504 & 0.485 & 0.481 & 0.458 \\
      Min-20\%        & 0.597 & 0.588 & 0.528 & 0.503 & 0.487 & 0.474 & 0.457 \\
      Min-30\%        & 0.604 & 0.595 & 0.539 & 0.510 & 0.491 & 0.477 & 0.460 \\
      Min-20\%++        & 0.562 & 0.579 & 0.519 & 0.481 & 0.457 & 0.440 & 0.418 \\
      Min-30\%++        & 0.601 & 0.610 & 0.565 & 0.517 & 0.488 & 0.473 & 0.449 \\
      Ref        & 0.597 & 0.579 & 0.567 & 0.588 & 0.589 & 0.603 & 0.602 \\
      Neighbor        & 0.571 & 0.560 & 0.547 & 0.524 & 0.513 & 0.499 & 0.499 \\
      CodeMI & 0.501 & 0.489 & 0.477 & 0.454 & 0.452 & 0.455 & 0.458 \\
      AdvPrompt & 0.587 & 0.610 & 0.606 & 0.589 & 0.583 & 0.573 & 0.568 \\
      ReCaLL        & \best{0.830} & \best{0.839} & \best{0.811} & \best{0.766} & \best{0.735} & \best{0.730} & \best{0.699} \\
      \bottomrule
    \end{tabular}
    \end{adjustbox}
  \end{minipage}
  
  \begin{minipage}[t]{0.47\textwidth}
    \centering
    \textbf{CodeLlama 13B}\par
    \vspace{2pt}
    \begin{adjustbox}{width=\linewidth}
    \begin{tabular}{l*{7}{S}}
      \toprule
      \multirow{1}{*}{Method} 
       & {Org} & {MT1} & {MT2a} & $\text{MT3}_{0.9}$ & $\text{MT3}_{0.7}$ & $\text{MT3}_{0.5}$ & {Hyb} \\
      \midrule
      PPL        & 0.653 & 0.649 & 0.585 & 0.550 & 0.525 & 0.516 & 0.501 \\
      Zlib        & 0.631 & 0.633 & 0.561 & 0.536 & 0.516 & 0.512 & 0.493 \\
      Min-20\%        & 0.645 & 0.640 & 0.572 & 0.537 & 0.518 & 0.505 & 0.492 \\
      Min-30\%        & 0.648 & 0.644 & 0.578 & 0.541 & 0.519 & 0.506 & 0.492 \\
      Min-20\%++        & 0.592 & 0.611 & 0.541 & 0.497 & 0.462 & 0.447 & 0.430 \\
      Min-30\%++        & 0.619 & 0.631 & 0.572 & 0.528 & 0.488 & 0.478 & 0.459 \\
      Ref        & 0.494 & 0.485 & 0.464 & 0.476 & 0.463 & 0.464 & 0.473 \\
      Neighbor        & 0.594 & 0.582 & 0.569 & 0.543 & 0.532 & 0.516 & 0.517 \\
      CodeMI & 0.531 & 0.518 & 0.503 & 0.479 & 0.475 & 0.475 & 0.483 \\
      AdvPrompt & 0.516 & 0.555 & 0.543 & 0.531 & 0.538 & 0.536 & 0.537 \\
      ReCaLL        & \best{0.813} & \best{0.795} & \best{0.765} & \best{0.696} & \best{0.651} & \best{0.634} & \best{0.612} \\
      \bottomrule
    \end{tabular}
    \end{adjustbox}
  \end{minipage}
  \hfill
  \begin{minipage}[t]{0.47\textwidth}
    \centering
    \textbf{DeepSeekCoder 33B}\par
    \vspace{2pt}
    \begin{adjustbox}{width=\linewidth}
    \begin{tabular}{l*{7}{S}}
      \toprule
      \multirow{1}{*}{Method} 
       & {Org} & {MT1} & {MT2a} & $\text{MT3}_{0.9}$ & $\text{MT3}_{0.7}$ & $\text{MT3}_{0.5}$ & {Hyb} \\
      \midrule
      PPL        & 0.634 & 0.617 & 0.558 & 0.522 & 0.502 & 0.490 & 0.473 \\
      Zlib        & 0.611 & 0.599 & 0.534 & 0.507 & 0.492 & 0.486 & 0.464 \\
      Min-20\%        & 0.625 & 0.610 & 0.545 & 0.508 & 0.493 & 0.476 & 0.462 \\
      Min-30\%        & 0.629 & 0.614 & 0.552 & 0.513 & 0.494 & 0.480 & 0.464 \\
      Min-20\%++        & 0.595 & 0.601 & 0.550 & 0.497 & 0.471 & 0.446 & 0.435 \\
      Min-30\%++        & 0.618 & 0.617 & 0.570 & 0.521 & 0.485 & 0.473 & 0.453 \\
      Ref        & 0.656 & 0.659 & 0.622 & 0.619 & 0.608 & 0.605 & 0.594 \\
      Neighbor        & 0.588 & 0.577 & 0.565 & 0.540 & 0.527 & 0.512 & 0.510 \\
      CodeMI & 0.513 & 0.501 & 0.487 & 0.465 & 0.462 & 0.461 & 0.468 \\
      AdvPrompt & 0.604 & 0.594 & 0.548 & 0.549 & 0.550 & 0.522 & 0.530 \\
      ReCaLL        & \best{0.806} & \best{0.800} & \best{0.786} & \best{0.721} & \best{0.681} & \best{0.666} & \best{0.642} \\
      \bottomrule
    \end{tabular}
    \end{adjustbox}
  \end{minipage}
  \hfill
  \\ [1em]

  \label{tab:test1_auc}
  \vspace{-0.1cm}
\end{table*}

\subsection{RQ2: Effectiveness under Mutation}
To address the research question, we evaluate the robustness of PDD methods against mutated member samples under three testing settings, referred to as Setting 1, Setting 2,  and Setting 3. 

\textbf{Setting 1: Member-only Mutation.}
Under this setting, we uniformly assume that the mutated membership data should still be regarded as membership data. Thus, we replace the membership dataset in \tool with its mutated version, while keeping the non-membership data unchanged. 
We separately apply \textbf{Mutation-T1} ($MT1$), \textbf{Mutation-T2} ($MT2$), \textbf{Mutation-T3} ($MT3$), and \textbf{Hybrid} ($Hyb$) to the membership data. Thus, we obtain multiple test sets: $\mathcal{D}_{s1}^{1}=\{MT1(\mathcal{D}_m), \mathcal{D}_{nm}\}$, $\mathcal{D}_{s1}^{2}=\{MT2(\mathcal{D}_m), \mathcal{D}_{nm}\}$, $\mathcal{D}_{s1}^{3}=\{MT3(\mathcal{D}_m), \mathcal{D}_{nm}\}$, etc.

As shown in Table~\ref{tab:test1_auc}, compared to their performance on the original \tool, most PDD methods degrade when the member data is mutated, while AdvPrompt-MIA exhibits model- and mutation-dependent fluctuations.
For {Mutation-T1}, which reformats the code style of member samples, only minor performance fluctuations are observed.
This suggests that whitespace and formatting characters, while handled differently by each method, do not play a decisive role in their predictions.
For {Mutation-T2a}, which randomly renames identifiers, most PDD methods experience a degradation in their AUC scores.
For example, the AUC for the Min-30\% method drops from 0.604 to 0.539 on StarCoder2-7B. This highlights a key vulnerability of existing PDD methods: an over-reliance on lexical tokens, such as variable names, which are not integral to code semantics. 
To further investigate the impact of lexical changes, we employ Mutation-T2b, which simulates a worst-case scenario. In this setup, all identifiers are replaced with random 8-character strings. As shown in Fig.~\ref{fig:2ato2c}, this obfuscation causes the performance of the methods included in this analysis to deteriorate further. Notably, the effectiveness of several methods degrades to a level approaching random guessing.


For Mutation-T3, we observe that methods relying solely on token generation probabilities, such as PPL, Zlib, Min-K\%, and Min-K\%++, exhibit further performance degradation. For instance, on StarCoder2-7B, the AUC for Min-30\%++ drops across the four mutation levels to 0.517, 0.488, 0.473, and 0.449, respectively. 
In contrast, reference-based methods like Ref and Neighbor demonstrate stable or even improved performance against Type-3 mutations. 
This indicates that methods computing a relative score against a reference set show greater robustness to statement-level substitutions.
Among all these PDD methods, ReCaLL remains the most effective method against $\text{Mutation-T3}_{(0.5)}$, achieving a state-of-the-art average AUC of 0.730 on StarCoder2-7B.

\begin{center}
    \includegraphics[width=\linewidth, keepaspectratio]{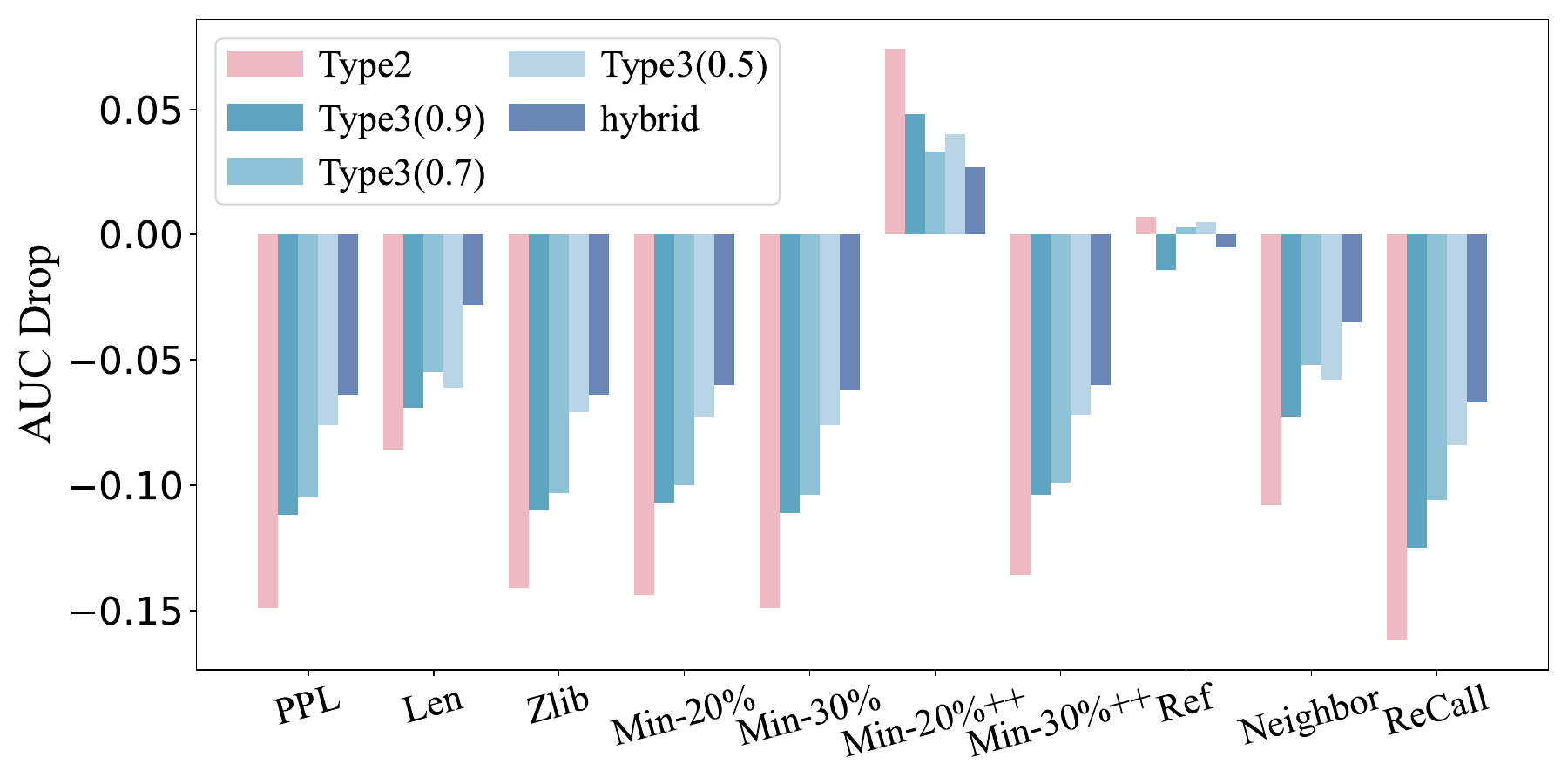}
    \captionof{figure}{The AUC drop of various methods when the Type-2 mutation strategy is changed from identifier renaming (Mutation-T2a) to random string replacement (Mutation-T2b).}
    \label{fig:2ato2c}
\end{center}

\textbf{Setting 2: Clone-Type-Aware Mutations.} 
Under Setting 2, we adopt a clone-type-aware assumption. Specifically, under a Type-$l$ clone definition, only membership data mutated up to level $l$ is still regarded as membership data. For instance, with $l{=}1$, samples mutated by {Mutation-T1} remain members, while those mutated by {Mutation-T2} or beyond are treated as non-members. 
This is because, under the Type-$1$ clone standard, samples mutated by {Mutation-T1} are regarded as identical to the original samples, whereas samples mutated by {Mutation-T2} are considered different. Meanwhile, under any code clone definition, non-members would remain non-members after any mutation strategy is applied. To ensure a balanced comparison between member and non-member sets, we apply the same levels of mutation to the non-member data as are permitted for the member data.
That is, the members include $\mathcal{D}_m$ and $ MT1(\mathcal{D}_m)$. The non-members include $MT2(\mathcal{D}_m)$, $MT3(\mathcal{D}_m)$, $\mathcal{D}_{nm}$ and $MT1(\mathcal{D}_{nm})$. 
Similarly, with $l{=}2$, samples mutated by {Mutation-T1}  and {Mutation-T2} remain members, while those mutated by {Mutation-T3} or beyond are treated as non-members. 
That is, the members include $\mathcal{D}_m$, $ MT1(\mathcal{D}_m)$, and $MT2(\mathcal{D}_m)$. The non-members include $MT3(\mathcal{D}_m)$, $\mathcal{D}_{nm}$, $MT1(\mathcal{D}_{nm})$, $MT2(\mathcal{D}_{nm})$. 
This setup reflects scenarios considering code clone definitions.


Table~\ref{tab:test8_auc} presents the results. 
Note that in addition to MT1, MT2, MT3, and Hyb, we also include the Verbatim type in the table, which refers to cases where any mutation applied to the members is treated as non-member data.
The overall performance trends are consistent with Setting 1, with most methods achieving an AUC around 0.6 and ReCaLL remaining the top performer with an AUC around 0.7. However, the performance degradation with increasing $K$ is less steep. For example, on StarCoder2-7B, while the AUC of PPL drops from 0.608 to 0.466 in Setting 1, its drop in Setting 2 is only from 0.654 to 0.611.

%

\begin{table*}[htbp]
  \centering
  \setlength{\tabcolsep}{4pt}
  \renewcommand{\arraystretch}{1.1}
  \vspace{-0.2cm}
\caption{AUC scores under Setting 2 (Clone-Type-Aware Mutations). The ``–'' indicates that no result is available due to the absence of a reference model for the 1.1B SantaCoder. Type 1, Type 2, and Type 3 represent different clone definitions, where $\text{Type 3}_{(0.9)}$ denotes a Type 3 clone identified using a similarity threshold of 0.9. For the Verbatim type, any mutation applied to the members is considered non-member data.}
  \begin{minipage}[t]{0.47\textwidth}
    \centering
    \textbf{SantaCoder 1.1B}\par
    \vspace{2pt}
    \begin{adjustbox}{width=\linewidth}
    \begin{tabular}{l*{6}{S}}
      \toprule
     \multirow{1}{*}{Clone Type} 
       & {\footnotesize Verbatim} & { Type 1} & { Type 2} & {\footnotesize $\text{Type 3}_{(0.9)}$} & {\footnotesize $\text{Type 3}_{(0.7)}$} & {\footnotesize $\text{Type 3}_{(0.5)}$}\\
       \midrule
      PPL        & 0.640 & 0.658 & 0.635 & 0.608 & 0.615 & 0.596 \\
      Zlib        & 0.617 & 0.639 & 0.603 & 0.583 & 0.605 & 0.592 \\
      Min-20\%        & 0.632 & 0.651 & 0.619 & 0.598 & 0.609 & 0.594 \\
      Min-30\%        & 0.638 & 0.658 & 0.629 & 0.606 & 0.614 & 0.598 \\
      Min-20\%++        & 0.609 & 0.638 & 0.625 & 0.598 & 0.580 & 0.574 \\
      Min-30\%++        & 0.634 & 0.661 & 0.650 & 0.620 & 0.608 & 0.595 \\
      Ref        & \text{-} & \text{-} & \text{-} & \text{-} & \text{-} & \text{-} \\
      Neighbor        & 0.570 & 0.587 & 0.591 & 0.591 & 0.573 & 0.559 \\
      CodeMI & 0.498 & 0.530 & 0.510 & 0.511 & 0.481 & 0.459 \\
      AdvPrompt & 0.592 & 0.529 & 0.577 & 0.577 & 0.576 & 0.565 \\
      ReCaLL        & \best{0.749} & \best{0.768} & \best{0.746} & \best{0.716} & \best{0.713} & \best{0.725} \\
      \bottomrule
    \end{tabular}
    \end{adjustbox}
  \end{minipage}
  \hfill
  \begin{minipage}[t]{0.47\textwidth}
    \centering
    \textbf{StarCoder2 7B}\par
    \vspace{2pt}
    \begin{adjustbox}{width=\linewidth}
    \begin{tabular}{l*{6}{S}}
      \toprule
      \multirow{1}{*}{Clone Type} 
       & {\footnotesize Verbatim} & { Type 1} & { Type 2} & {\footnotesize $\text{Type 3}_{(0.9)}$} & {\footnotesize $\text{Type 3}_{(0.7)}$} & {\footnotesize $\text{Type 3}_{(0.5)}$}\\
      \midrule
      PPL        & 0.654 & 0.666 & 0.639 & 0.610 & 0.621 & 0.611 \\
      Zlib        & 0.630 & 0.646 & 0.609 & 0.586 & 0.611 & 0.609 \\
      Min-20\%        & 0.645 & 0.656 & 0.624 & 0.599 & 0.614 & 0.609 \\
      Min-30\%        & 0.652 & 0.666 & 0.636 & 0.607 & 0.621 & 0.613 \\
      Min-20\%++        & 0.617 & 0.647 & 0.642 & 0.605 & 0.591 & 0.603 \\
      Min-30\%++        & 0.648 & 0.681 & 0.671 & 0.629 & 0.625 & 0.622 \\
      Ref        & 0.569 & 0.525 & 0.525 & 0.530 & 0.523 & 0.551 \\
      Neighbor        & 0.569 & 0.588 & 0.590 & 0.587 & 0.573 & 0.561 \\
      CodeMI & 0.498 & 0.527 & 0.514 & 0.509 & 0.483 & 0.464 \\
      AdvPrompt & 0.595 & 0.541 & 0.530 & 0.570 & 0.566 & 0.566 \\
      ReCaLL        & \best{0.769} & \best{0.779} & \best{0.760} & \best{0.737} & \best{0.738} & \best{0.735} \\
      \bottomrule
    \end{tabular}
    \end{adjustbox}
  \end{minipage}
  \begin{minipage}[t]{0.47\textwidth}
    \centering
    \textbf{CodeLlama 13B}\par
    \vspace{2pt}
    \begin{adjustbox}{width=\linewidth}
    \begin{tabular}{l*{6}{S}}
      \toprule
      \multirow{1}{*}{Clone Type} 
       & {\footnotesize Verbatim} & {Type 1} & {Type 2} & {\footnotesize $\text{Type 3}_{(0.9)}$} & {\footnotesize $\text{Type 3}_{(0.7)}$} & {\footnotesize $\text{Type 3}_{(0.5)}$}\\
      \midrule
      PPL        & 0.685 & 0.701 & 0.658 & 0.636 & 0.645 & 0.646 \\
      Zlib        & 0.660 & 0.682 & 0.627 & 0.609 & 0.633 & 0.640 \\
      Min-20\%        & 0.683 & 0.702 & 0.653 & 0.633 & 0.641 & 0.646 \\
      Min-30\%        & 0.685 & 0.705 & 0.659 & 0.637 & 0.645 & 0.647 \\
      Min-20\%++        & 0.641 & 0.674 & 0.661 & 0.612 & 0.599 & 0.614 \\
      Min-30\%++        & 0.659 & 0.691 & 0.672 & 0.634 & 0.628 & 0.627 \\
      Ref        & 0.509 & 0.497 & 0.488 & 0.488 & 0.512 & 0.527 \\
      Neighbor        & 0.589 & 0.609 & 0.610 & 0.609 & 0.590 & 0.578 \\
      CodeMI & 0.532 & 0.566 & 0.557 & 0.540 & 0.510 & 0.484 \\
      AdvPrompt & 0.520 & 0.538 & 0.523 & 0.541 & 0.547 & 0.544 \\
      ReCaLL        & \best{0.799} & \best{0.797} & \best{0.787} & \best{0.745} & \best{0.736} & \best{0.737} \\
      \bottomrule
    \end{tabular}
    \end{adjustbox}
  \end{minipage}
  \hfill
  \begin{minipage}[t]{0.47\textwidth}
    \centering
    \textbf{DeepSeekCoder 33B}\par
    \vspace{2pt}
    \begin{adjustbox}{width=\linewidth}
    \begin{tabular}{l*{6}{S}}
      \toprule
      \multirow{1}{*}{Clone Type} 
       & {\footnotesize Verbatim} & {Type 1} & {Type 2} & {\footnotesize $\text{Type 3}_{(0.9)}$} & {\footnotesize $\text{Type 3}_{(0.7)}$} & {\footnotesize $\text{Type 3}_{(0.5)}$}\\
      \midrule
      PPL        & 0.675 & 0.687 & 0.647 & 0.625 & 0.632 & 0.633 \\
      Zlib        & 0.649 & 0.666 & 0.615 & 0.599 & 0.619 & 0.629 \\
      Min-20\%        & 0.671 & 0.688 & 0.643 & 0.623 & 0.630 & 0.633 \\
      Min-30\%        & 0.674 & 0.691 & 0.648 & 0.626 & 0.633 & 0.634 \\
      Min-20\%++        & 0.642 & 0.673 & 0.668 & 0.619 & 0.611 & 0.623 \\
      Min-30\%++        & 0.659 & 0.685 & 0.673 & 0.631 & 0.628 & 0.625 \\
      Ref        & 0.607 & 0.607 & 0.566 & 0.573 & 0.582 & 0.606 \\
      Neighbor        & 0.583 & 0.601 & 0.606 & 0.607 & 0.588 & 0.575 \\
      CodeMI & 0.512 & 0.548 & 0.545 & 0.528 & 0.496 & 0.478 \\
      AdvPrompt & 0.602 & 0.555 & 0.538 & 0.548 & 0.534 & 0.535 \\
      ReCaLL        & \best{0.760} & \best{0.779} & \best{0.782} & \best{0.741} & \best{0.733} & \best{0.703} \\
      \bottomrule
    \end{tabular}
    \end{adjustbox}
  \end{minipage}
  \hfill
  \\ [1em]

\vspace{-0.6cm}
  \label{tab:test8_auc}
\end{table*}

We observe that for most methods, the AUC score improves from the Verbatim type to $l=1$. For instance, on StarCoder2-7B, Neighbor improves from 0.569 to 0.588, while Min-20\%++ improves from 0.617 to 0.647.
Our analysis is as follows: at $l=0$, samples mutated by {Mutation-T1} on original members ($MT1(\mathcal{D}_m)$) are included in the non-member set. Such data is inherently hard to distinguish from the original member samples. At $l=1$, these samples $MT1(\mathcal{D}_m)$ are treated as members, leading to higher detection performance.


For $l \ge 2$, most methods exhibit a declining performance trend as $K$ increases, although AdvPrompt-MIA fluctuates across clone types and models. For instance, ReCaLL drops from 0.782 to 0.741 on DeepSeekCoder 33B.
This highlights the challenge of maintaining detection robustness under high mutation levels.

Overall, the worst-case AUC for most methods under Setting 2 is approximately 0.6, while for ReCaLL it is around 0.7. This test likely provides a better estimate of worst-case performance by considering a more comprehensive set of scenarios.




To investigate the impact of each specific code mutation strategy, we apply a single, isolated transformation to both Python member and non-member sets and observe the resulting AUC scores.

\textbf{Setting 3: Isolated Mutation Impact.}
Under this setting, we apply a specific level of mutation synchronously to both original member and non-member samples to construct the respective test sets. Thus, we have multiple test sets: $\mathcal{D}_{s3}^{1}=\{MT1(\mathcal{D}_m), MT1(\mathcal{D}_{nm})\}$, $\mathcal{D}_{s3}^{2}=\{MT2(\mathcal{D}_m), MT2(\mathcal{D}_{nm})\}$, etc.

\begin{figure}[htb]
    \centering
    \includegraphics[width=\linewidth, keepaspectratio]{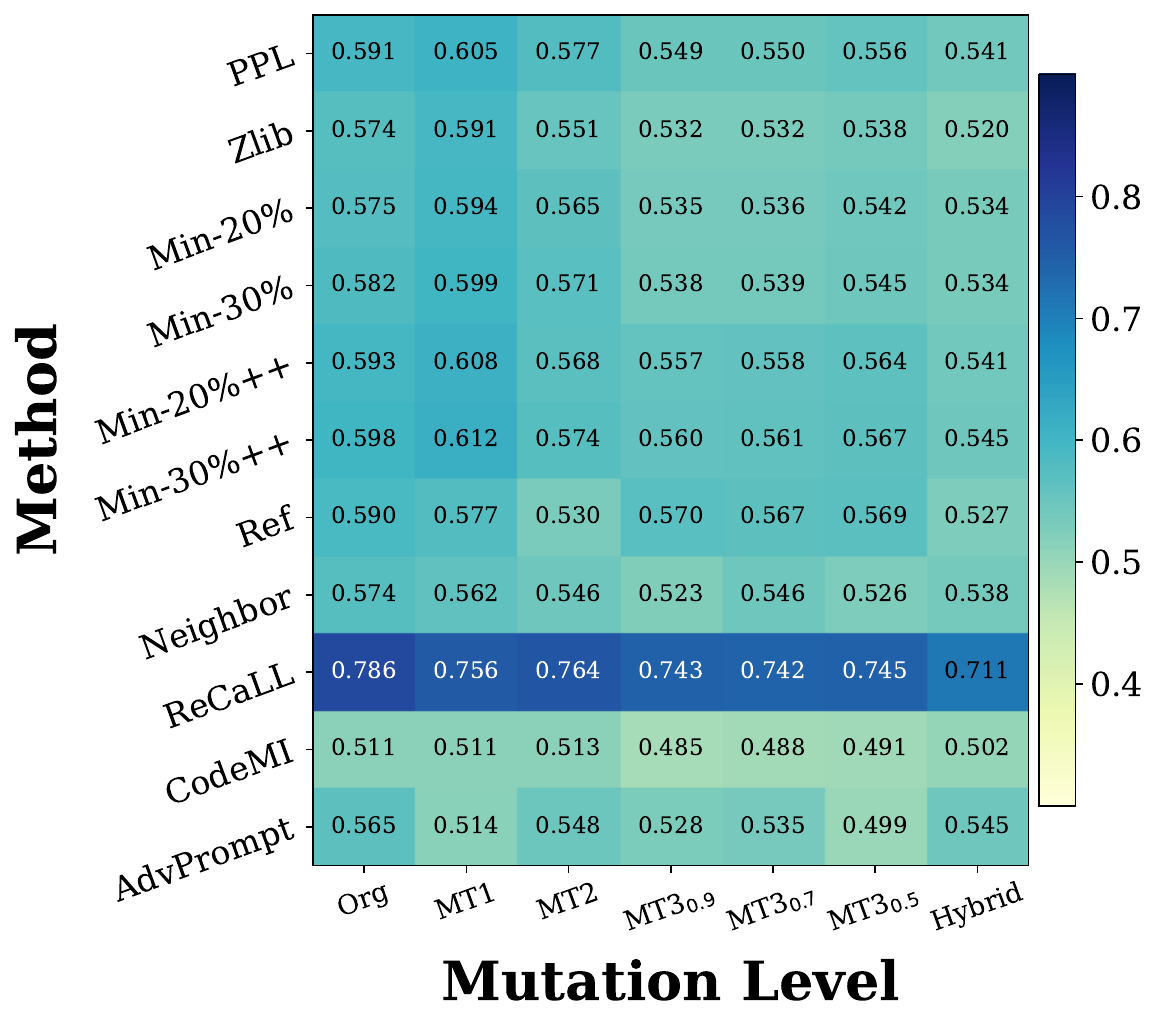}
    \captionof{figure}{AUC Scores under Setting 3 (Isolated Mutation Impact)}
    \label{fig:heatmap1}
\end{figure}

Subsequently, we evaluated the performance of all PDD methods on the test sets $\mathcal{D}^l_{s3}$ corresponding to different mutation levels on the StarCoder2-7B model. This setting allows us to demonstrate the impact of applying each mutation type individually to the test set.
As shown in Fig.~\ref{fig:heatmap1}, methods generally maintain an AUC around 0.6, with ReCaLL achieving nearly 0.8. 
For Mutation-T1 and Mutation-T2, token-based methods like PPL, Zlib, and Min-K\% remain relatively stable (\eg, PPL's AUCs are 0.591, 0.605, and 0.577 for the first three levels), demonstrating resilience to changes in formatting and identifiers.
Conversely, Ref and ReCaLL show a slight AUC drop. For ReCaLL, this may occur because the model first learns the formatting style from the non-member prefix and then encounters a differently styled target snippet, leading to prediction errors that affect the relative loss calculation for both members and non-members. CodeMI and AdvPrompt-MIA obtain AUCs of 0.485--0.513 and 0.499--0.565 across the mutation levels, respectively.

For Mutation-T3, the performance of token-based methods degrades, revealing their sensitivity to semantically equivalent statement substitutions. For example, Min-30\%++'s AUC drops from 0.574 to 0.560. This suggests that while robust to lexical changes, these methods struggle with syntactic paraphrasing, as novel statement structures may generate low-probability tokens that are incorrectly flagged as non-member features. In contrast, Ref and ReCaLL exhibit stability (\eg, Ref maintains an AUC of 0.570, comparable to its score of 0.590 on the original data). Their reliance on relative comparisons—either against a reference model or a prefixed context—mitigates the direct impact of sample loss inflation caused by mutations, showcasing the promise of reference-based approaches for mutated code.

For the most complex mutation (\textbf{Hybrid}), the performance of all methods drops to around 0.55 AUC, with even ReCaLL falling from 0.786 to 0.711, indicating that sophisticated Type-3 clones remain a significant challenge. Neighbor remains near-random under this setting (0.538 on StarCoder2-7B), even with code-aware neighborhood generation, showing that the neighborhood signal is not robust to compound mutations. CodeMI and AdvPrompt-MIA achieve AUC scores of 0.502 and 0.545, respectively.

\begin{tcolorbox}[size=title]
	{\textbf{Answer to RQ2: Under different settings, mutations lead to a degradation in detection performance. ReCaLL consistently achieves the best results. This further highlights that robust PDD methods are needed for code.}}
\end{tcolorbox}

\subsection{RQ3: Impact of Code Length}
\begin{figure}[htb]
    \centering
    \includegraphics[width=\linewidth, keepaspectratio]{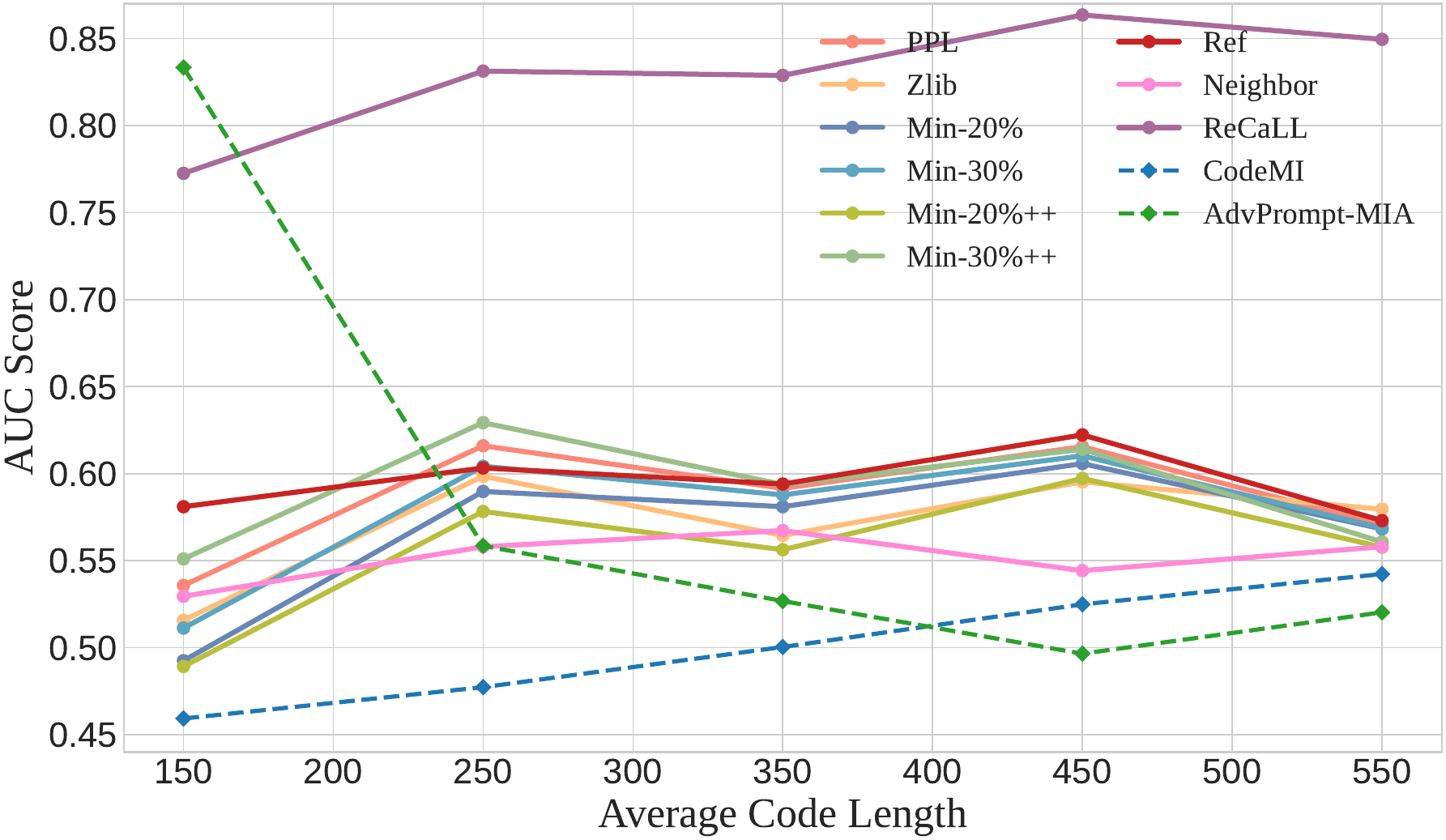}
    \captionof{figure}{AUC Scores as Code Length Increases}
    \vspace{-0.5cm}
    \label{fig:test0_length}
\end{figure}

To answer this research question, we partitioned \tool into five subsets based on code length, creating distinct test sets for functions with code lengths in the ranges of 100-200, 200-300, 300-400, 400-500, and 500-600. We then selected the StarCoder2-7B model and evaluated the performance of nine PDD methods on each of these five subsets.
Fig.~\ref{fig:test0_length} illustrates the relationship between average code length and the resulting AUC scores.

{We observe a method-dependent relationship between code length and detection performance. As the average code length increases from 150 to 250, most established methods improve; for example, PPL’s AUC rises from 0.536 to 0.615, indicating that detection is more challenging for very short code. CodeMI also increases from 0.459 to 0.477, whereas AdvPrompt-MIA decreases from 0.833 to 0.559.}

However, increased code length does not consistently lead to better performance. For instance, the AUC of the Min-k\%++ method drops from 0.635 to 0.588 as the average sample length increases from 250 to 350. This inconsistency contrasts with findings on natural language data, where performance generally continues to improve with longer text~\cite{zhang2025pretrainingdatadetectionlarge}.

We additionally evaluate CodeLlama-13B on the same five code-length subsets and observe the same pattern: increasing code length does not consistently improve detection. Full results are provided on our project website~\cite{website}.

\begin{tcolorbox}[size=title]
	{\textbf{Answer to RQ3: Increasing code length from 150 to 250 improves the performance of most established methods, but the effect is method-dependent and further increases do not consistently yield better results. }}
\end{tcolorbox}

\subsection{RQ4: Consistency Across Different Programming Languages}
\begin{table*}[htbp]
  \centering
  \setlength{\tabcolsep}{4pt}
  \renewcommand{\arraystretch}{1.1}
  \caption{AUC scores for different methods across C++ and Java languages for Setting 1 and Setting 2. Higher is better; best results are highlighted in \textbf{bold}. }
  
  
  \vspace{-0.2cm}
  \begin{minipage}[t]{0.48\textwidth}
    \centering
    \textbf{C++ (Setting 1)}\par
    \vspace{2pt}
    \begin{adjustbox}{width=\linewidth}
    \begin{tabular}{l*{7}{S}}
      \toprule
     \multirow{1}{*}{Method} 
       & {Org} & {MT1} & {MT2a} & $\text{MT3}_{0.9}$ & $\text{MT3}_{0.7}$ & $\text{MT3}_{0.5}$ & {Hyb} \\
      \midrule
      PPL         & 0.541 & 0.532 & 0.488 & 0.499 & 0.492 & 0.467 & 0.446 \\
      Zlib        & 0.545 & 0.536 & 0.516 & 0.522 & 0.520 & 0.506 & 0.488 \\
      Min-20\%    & 0.529 & 0.520 & 0.477 & 0.490 & 0.478 & 0.450 & 0.430 \\
      Min-30\%    & 0.532 & 0.523 & 0.479 & 0.493 & 0.482 & 0.455 & 0.435 \\
      Min-20\%++  & 0.520 & 0.512 & 0.473 & 0.507 & 0.503 & 0.483 & 0.473 \\
      Min-30\%++  & 0.540 & 0.532 & 0.493 & 0.514 & 0.513 & 0.497 & 0.479 \\
      Ref         & 0.490 & 0.480 & 0.484 & 0.468 & 0.472 & 0.447 & 0.469 \\
      Neighbor & 0.441 & 0.476 & 0.462 & 0.464 & 0.448 & 0.429 & 0.427 \\
      CodeMI & 0.427 & 0.416 & 0.419 & 0.431 & 0.440 & 0.431 & 0.438 \\
      AdvPrompt & 0.513 & 0.537 & 0.548 & 0.566 & 0.571 & 0.548 & 0.548 \\
      ReCaLL      & \best{0.857} & \best{0.837} & \best{0.849} & \best{0.845} & \best{0.840} & \best{0.831} & \best{0.824} \\
      \bottomrule
    \end{tabular}
    \end{adjustbox}
  \end{minipage}
  \hfill
  \begin{minipage}[t]{0.48\textwidth}
    \centering
    \textbf{Java (Setting 1)}\par
    \vspace{2pt}
    \begin{adjustbox}{width=\linewidth}
    \begin{tabular}{l*{7}{S}}
        \toprule
        \multirow{1}{*}{Method} 
       & {Org} & {MT1} & {MT2a} & $\text{MT3}_{0.9}$ & $\text{MT3}_{0.7}$ & $\text{MT3}_{0.5}$ & {Hyb} \\
        \midrule
        PPL         & 0.660 & 0.634 & 0.587 & 0.605 & 0.583 & 0.589 & 0.584 \\
        Zlib        & 0.671 & 0.639 & 0.608 & 0.621 & 0.604 & 0.610 & 0.613 \\
        Min-20\%    & 0.666 & 0.641 & 0.589 & 0.610 & 0.582 & 0.582 & 0.576 \\
        Min-30\%    & 0.664 & 0.639 & 0.589 & 0.609 & 0.584 & 0.586 & 0.579 \\
        Min-20\%++  & 0.588 & 0.579 & 0.544 & 0.568 & 0.553 & 0.558 & 0.565 \\
        Min-30\%++  & 0.630 & 0.611 & 0.572 & 0.591 & 0.576 & 0.595 & 0.588 \\
        Ref         & 0.543 & 0.540 & 0.528 & 0.530 & 0.535 & 0.545 & 0.534 \\
        Neighbor & 0.500 & 0.423 & 0.403 & 0.425 & 0.415 & 0.421 & 0.420 \\
        CodeMI & 0.525 & 0.490 & 0.487 & 0.506 & 0.505 & 0.524 & 0.539 \\
        AdvPrompt & 0.611 & 0.611 & 0.616 & 0.641 & 0.636 & 0.642 & 0.649 \\
        ReCaLL      & \best{0.842} & \best{0.809} & \best{0.793} & \best{0.803} & \best{0.802} & \best{0.801} & \best{0.801} \\
        \bottomrule
    \end{tabular}
    \end{adjustbox}
  \end{minipage}
  \\ [1em]
  \begin{minipage}[t]{0.48\textwidth}
    \centering
    \textbf{C++ (Setting 2)}\par
    \vspace{2pt}
    \begin{adjustbox}{width=\linewidth}
    \begin{tabular}{l*{6}{S}}
        \toprule
        \multirow{1}{*}{Clone Type} 
       & {\footnotesize Verbatim} & {Type 1} & {Type 2} & {\footnotesize $\text{Type 3}_{(0.9)}$} & {\footnotesize $\text{Type 3}_{(0.7)}$} & {\footnotesize $\text{Type 3}_{(0.5)}$}\\
        \midrule
        PPL         & 0.558 & 0.572 & 0.530 & 0.584 & 0.552 & 0.581 \\
        Zlib        & 0.542 & 0.549 & 0.521 & 0.559 & 0.534 & 0.546 \\
        Min-20\%    & 0.556 & 0.577 & 0.534 & 0.581 & 0.553 & 0.580 \\
        Min-30\%    & 0.556 & 0.575 & 0.531 & 0.581 & 0.552 & 0.582 \\
        Min-20\%++  & 0.526 & 0.524 & 0.502 & 0.529 & 0.516 & 0.543 \\
        Min-30\%++  & 0.542 & 0.541 & 0.507 & 0.554 & 0.529 & 0.562 \\
        Ref         & 0.519 & 0.522 & 0.506 & 0.511 & 0.508 & 0.490 \\
        Neighbor & 0.482 & 0.526 & 0.513 & 0.510 & 0.520 & 0.480 \\
        CodeMI & 0.410 & 0.441 & 0.415 & 0.466 & 0.419 & 0.412 \\
        AdvPrompt & 0.507 & 0.455 & 0.431 & 0.527 & 0.502 & 0.514 \\
        ReCaLL      & \best{0.719} & \best{0.708} & \best{0.689} & \best{0.728} & \best{0.694} & \best{0.713} \\
        \bottomrule
    \end{tabular}
    \end{adjustbox}
  \end{minipage}
  \hfill
  \begin{minipage}[t]{0.48\textwidth}
    \centering
    \textbf{Java (Setting 2)}\par
    \vspace{2pt}
    \begin{adjustbox}{width=\linewidth}
    \begin{tabular}{l*{6}{S}}
        \toprule
        \multirow{1}{*}{Clone Type} 
       & {\footnotesize Verbatim} & {Type 1} & {Type 2} & {\footnotesize $\text{Type 3}_{(0.9)}$} & {\footnotesize $\text{Type 3}_{(0.7)}$} & {\footnotesize $\text{Type 3}_{(0.5)}$}\\
        \midrule
        PPL         & 0.624 & 0.605 & 0.602 & 0.606 & 0.620 & 0.630 \\
        Zlib        & 0.623 & 0.598 & 0.600 & 0.596 & 0.615 & 0.626 \\
        Min-20\%    & 0.632 & 0.616 & 0.612 & 0.619 & 0.626 & 0.635 \\
        Min-30\%    & 0.629 & 0.612 & 0.608 & 0.614 & 0.626 & 0.634 \\
        Min-20\%++  & 0.553 & 0.540 & 0.517 & 0.540 & 0.548 & 0.583 \\
        Min-30\%++  & 0.587 & 0.568 & 0.553 & 0.559 & 0.594 & 0.612 \\
        Ref         & 0.529 & 0.516 & 0.515 & 0.495 & 0.502 & 0.538 \\
        Neighbor & 0.542 & 0.504 & 0.487 & 0.507 & 0.505 & 0.512 \\
        CodeMI & 0.465 & 0.463 & 0.459 & 0.482 & 0.465 & 0.463 \\
        AdvPrompt & 0.562 & 0.477 & 0.475 & 0.543 & 0.548 & 0.565 \\
        ReCaLL      & \best{0.724} & \best{0.674} & \best{0.664} & \best{0.683} & \best{0.701} & \best{0.713} \\
        \bottomrule
    \end{tabular}
    \end{adjustbox}
  \end{minipage}

  \label{tab:pl_auc}
\end{table*}

Here, we replicate the Setting 1 and the Setting 2 on the StarCoder2-7B model for C++ and Java, in order to examine whether the conclusions remain consistent.

Table~\ref{tab:pl_auc} shows the results on C++ and Java. The "Org" column in the result table for Setting 1 shows the performance of all PDD methods on the original \tool, as no mutation is applied to the test samples in this setup. The performance trends after mutation observed in C++ and Java are largely consistent in Python. Under Setting 1, performance generally degrades as the mutation level increases, whereas under Setting 2, performance is more stable across levels.

The specific performance varies by language: On \tool, the average AUC score across all methods on C++ is 0.54, which is slightly lower than on Python (averaging around 0.59), while scores on Java are slightly higher (reaching 0.63). For the other parts of Setting 1 and Setting 2, the relative order of AUC scores remains consistent. For instance, in Setting 1, the ranking is C++ $<$ Python $<$ Java, while in Setting 2, Python and Java are comparable and both outperform C++. This variation may be attributed to differences in language syntax and verbosity. For the two code-specific methods, CodeMI obtains AUC scores of 0.410--0.539 and AdvPrompt-MIA obtains 0.431--0.649. ReCaLL retains the highest AUC across both settings.

Despite these variations, ReCaLL remains the top-performing method, achieving AUCs above 0.8 for C++ and Java under Setting 1 and around 0.7 under Setting 2.

We additionally evaluate CodeLlama-13B on C++ and Java under Settings 1 and 2. ReCaLL achieves the highest AUC in all evaluated conditions, while its AUC decreases under member-only Hybrid mutation in both languages. Full results are provided on our project website~\cite{website}.

\begin{tcolorbox}[size=title]
	{\textbf{Answer to RQ4: On the evaluated models, the conclusion is consistent across the three evaluated programming languages: ReCaLL consistently achieves the best performance, while the robustness of the evaluated methods still requires improvement.}}
\end{tcolorbox}

\subsection{RQ5: Sensitivity to Temporal Distribution Shift}
To answer RQ5, we retain the original member data, which provide high-confidence member labels based on existing pretraining corpora. We then apply the original non-member sampling principles to two later temporal windows. Following the construction of \tool, we collect functions from highly starred GitHub repositories and recent commits. The \textit{Late-2025} non-member data covers repositories created from May to December 2025 and commits from June to December 2025. The \textit{2026} non-member data covers repositories and commits from January to July 2026. We sample a fixed set of 600 functions from the original member data and pair it with 600 non-member functions from each temporal window, forming the Late-2025 and 2026 evaluation sets. We evaluate the original, unmutated code using the same eight CodeLLMs and the methods reported in Table~\ref{tab:temporal_rq5}.

\begin{table*}[ht]
\centering
\scriptsize
\caption{AUC comparison on original code under different non-member collection periods. Each evaluation set contains 600 member and 600 non-member samples. The Original row uses 600 randomly sampled non-members from the original benchmark, while the Late-2025 and 2026 rows use non-members from the corresponding later periods. Ref is averaged over the four models for which a reference model is available.}
\begin{adjustbox}{width=\textwidth}
{\begin{tabular}{l*{9}{c}}
\toprule
Non-member set & PPL & Zlib & Min-20\% & Min-30\% & Min-20\%++ & Min-30\%++ & Ref & Neighbor & ReCaLL \\
\midrule
Original & 0.578 & 0.545 & 0.563 & 0.571 & 0.581 & 0.585 & 0.590 & 0.574 & 0.786 \\
Late-2025 & 0.629 & 0.587 & 0.601 & 0.613 & 0.646 & 0.666 & 0.566 & 0.567 & 0.679 \\
2026 & 0.738 & 0.649 & 0.730 & 0.737 & 0.754 & 0.756 & 0.542 & 0.564 & 0.730 \\
\bottomrule
\end{tabular}}
\end{adjustbox}
\label{tab:temporal_rq5}
\end{table*}

Table~\ref{tab:temporal_rq5} reports the average AUC across the evaluated models. Following the grouping in RQ1, direct-score methods (PPL, zlib, Min-K\%, and Min-K\%++) aggregate or normalize the target model's token likelihoods for the candidate code. Their average AUCs increase monotonically from Original to Late-2025 and 2026. For example, PPL increases from 0.578 to 0.629 and 0.738, and Min-30\% increases from 0.571 to 0.737. Comparison-based methods construct relative scores using a reference model (Ref), generated neighbors (Neighbor), or non-member prefix context (ReCaLL). These methods show no common direction of change: Ref decreases from 0.590 to 0.542, Neighbor stays close to its original level, and ReCaLL first decreases and then recovers (0.786, 0.679, and 0.730).

A possible explanation for the direct-score trend is that newer code contains APIs, identifiers, and usage patterns that are less familiar to the evaluated models. Such code may receive lower token likelihoods, increasing its separation from the fixed member set and therefore raising AUC. PPL captures this change through the average log-likelihood, while Min-K\% focuses on the least likely tokens, which may include unfamiliar code elements. The increases for zlib and Min-K\%++ suggest that this sensitivity persists after calibration.

For comparison-based methods, we consider the relative response of the quantities being compared to explain their different temporal trends. In Ref, newer code may lower likelihoods under both the target and reference models, so subtraction can offset their shared response to temporal novelty. In Neighbor, the candidate and its generated neighbors share much of their code, allowing similar likelihood changes to cancel in the relative score. ReCaLL instead measures the change induced by a non-member prefix; its response may depend on how closely the prefix matches the code in each temporal window.

To our knowledge, this is the first study to identify these contrasting temporal trends between direct-score and comparison-based methods in function-level PDD for CodeLLMs. The choice of non-member window also changes their relative ranking: ReCaLL has a higher average AUC than Min-30\%++ on Original (0.786 versus 0.585), while Min-30\%++ has a higher average AUC on 2026 (0.756 versus 0.730). These findings alert future PDD studies to report the temporal provenance and collection windows of member and non-member data, compare methods under consistent temporal settings, and account for temporal sensitivity when interpreting detection performance.

\begin{tcolorbox}[size=title]
\textbf{Answer to RQ5:} Across the three evaluated non-member windows, direct-score methods show monotonically increasing average AUCs as the non-member data become more recent, while comparison-based methods show no common direction of change. Temporal windows affect both detection performance and method rankings and should be explicitly considered in PDD evaluation.
\end{tcolorbox}



\begin{center}
\includegraphics[width=\linewidth, keepaspectratio]{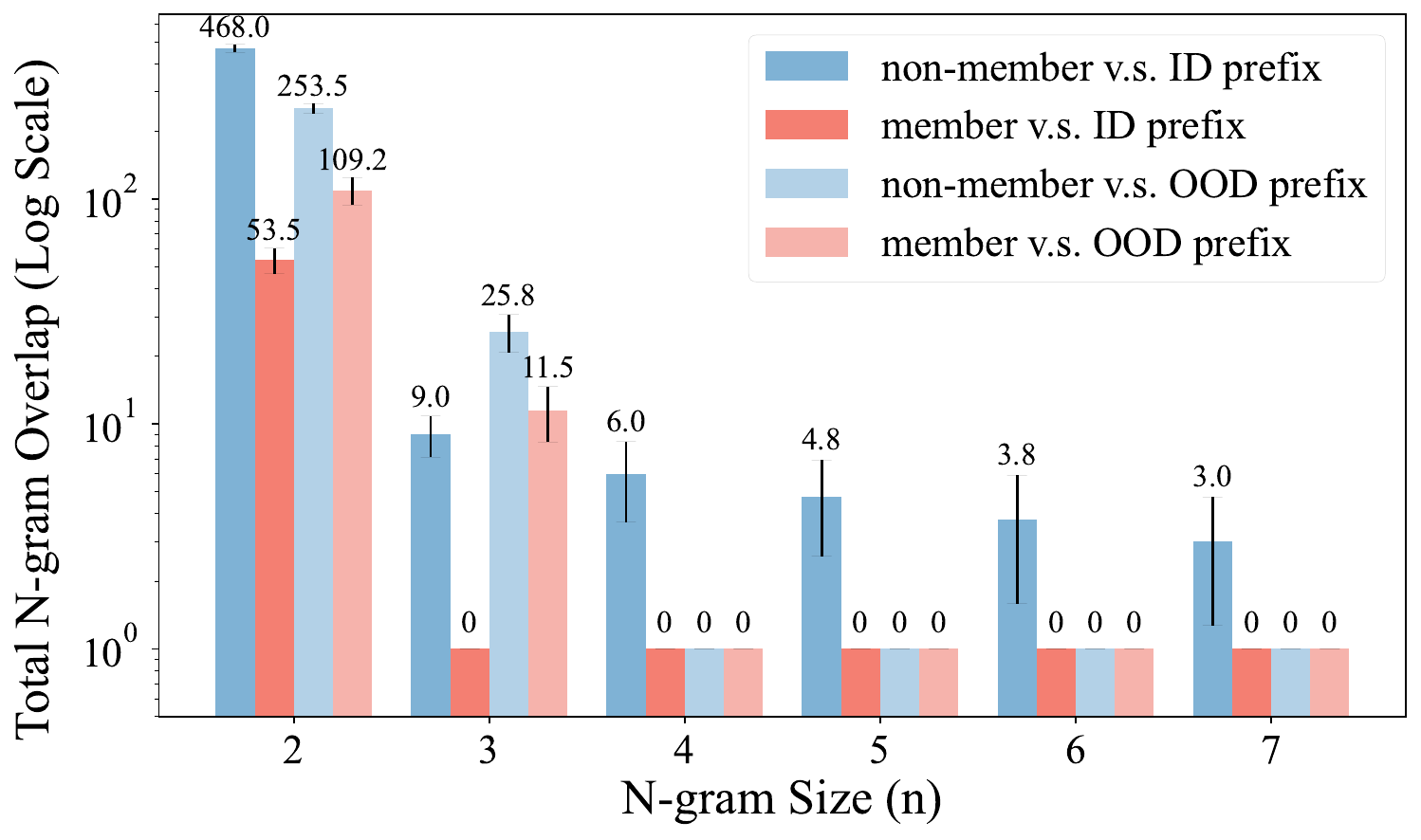}
    \captionof{figure}{Ngram overlap (in log scale) between prefix in ReCall and member \& nonmember dataset. Member v.s. ID prefix represents the overlap between the prefix sampled from the same distribution as the test set and the non-member dataset.}
    \label{fig:ngram}
\end{center}

\subsection{Discussion}

\textbf{A further investigation of ReCaLL.}
Prefix selection plays a crucial role in ReCaLL's success. To better understand this, we conduct a detailed analysis.
Specifically, we analyzed the n-gram overlap between the 12 prefixes  and \tool using the \texttt{StarCoder2} series of models. The result is shown in Fig.~\ref{fig:ngram}, which plots the n-gram overlap score against n, the length of the n-gram.
As the figure shows, the prefixes exhibit a markedly higher n-gram overlap with the non-member dataset than with the member dataset. Critically, for $n\ge3$, the overlap with the member data drops to zero, while a non-trivial overlap with the non-member data persists. This suggests that the prefixes, being drawn from the same non-member distribution, share specific features relevant to the non-member test data. This shared context likely causes the substantial loss reduction for non-member snippets when the prefix is applied, thereby boosting classification performance.

Note that in the previous setting, the prefixes are selected from $\mathcal{D}_{nm}$, which is drawn from the same distribution as the non-member set used during testing (as introduced in Section 4.2.5), and can therefore be regarded as in-distribution prefixes (\ie, ID prefixes).
To examine whether ReCaLL transfers across prefix sources, we experimented by selecting prefixes from out-of-distribution non-members instead (\ie, OOD prefixes). Specifically, we selected alternative GitHub repositories that also meet the previously described criteria and sampled prefixes from them. We observed that the n-gram overlap was reduced; specifically, for $n\ge4$, the overlap for both member and non-member sets dropped to zero. Although the overlap at larger $n$ is reduced, short n-gram overlaps can still reflect the shared temporal construction of the prefix and non-member sets.
With the OOD prefixes, we find that ReCaLL still achieves competitive performance. For example, on the \tool dataset, the AUC scores for ReCaLL with ID and OOD prefixes were 0.830 and 0.829, respectively, on the StarCoder2-7B model. These results show that ReCaLL remains effective when prefixes are drawn from alternative repositories, while this experiment does not isolate the effect of temporal distribution shift.

\section{Reverse Code Mutation for Pretraining Data Detection}

Previous RQs reveal a key challenge for PDD on code: although methods like ReCaLL can distinguish member and non-member samples on the original \tool, their performance drops substantially once the member code is mutated. In particular, reformatting, lexical renaming, and statement-level rewrites introduce a noticeable distribution shift between the original member samples seen during pretraining and the mutated versions. As a result, existing PDD methods may fail to recognize the memorized patterns and logic in code that were seen during pretraining, and instead become distracted by superficial changes in formatting, identifier surface forms, or semantically equivalent syntactic variants, leading to degradation in detection performance.

To address this challenge, we design \textbf{RCM} (Reverse Code Mutation), which consists of a \textit{reversion module} to further enhance the robustness of existing PDD methods under code mutations. The core idea is to reverse mutation artifacts in code via a series of normalization methods before applying PDD methods. If mutation-induced perturbations are removed in advance, the detector can operate on a code that is closer to the original form, thereby restoring the capability to distinguish member and non-member samples. As a preprocessing module, RCM is broadly compatible with existing PDD methods and can be applied to different detectors such as ReCaLL.

\begin{table*}[!t]
\centering
\caption{AUC scores under Setting 1 on a fixed subset of 300 member and 300 non-member functions. RCM-base retains the Type-1/Type-2 reversion stage and omits the Stage-2 rule-based reversion for Type-3 mutations.}
\label{tab:rq5_setting_1}
\resizebox{\textwidth}{!}{%
\begin{tabular}{lccccccccc}
\hline
Model & Method & Condition & Org & MT1 & MT2a & MT3$_{0.9}$ & MT3$_{0.7}$ & MT3$_{0.5}$ & Hyb \\
\hline
\multirow{6}{*}{CodeLlama-7B} & \multirow{3}{*}{PPL} & w/o RCM & 0.639 & 0.626 & 0.532 & 0.563 & 0.543 & 0.536 & 0.485 \\
 & & RCM-base & 0.635 & 0.635 & 0.603 & 0.623 & 0.621 & 0.619 & 0.590 \\
 & & RCM & 0.634 & 0.634 & 0.602 & 0.622 & 0.620 & 0.619 & 0.589 \\
\cline{2-10}
 & \multirow{3}{*}{ReCaLL} & w/o RCM & 0.811 & 0.768 & 0.732 & 0.571 & 0.569 & 0.569 & 0.508 \\
 & & RCM-base & 0.760 & 0.760 & 0.718 & 0.619 & 0.618 & 0.618 & 0.573 \\
 & & RCM & 0.760 & 0.760 & 0.719 & 0.619 & 0.618 & 0.618 & 0.573 \\
\midrule
\multirow{6}{*}{DeepSeekCoder-1.3B} & \multirow{3}{*}{PPL} & w/o RCM & 0.561 & 0.543 & 0.494 & 0.494 & 0.472 & 0.466 & 0.441 \\
 & & RCM-base & 0.599 & 0.599 & 0.579 & 0.584 & 0.583 & 0.582 & 0.563 \\
 & & RCM & 0.598 & 0.598 & 0.579 & 0.584 & 0.582 & 0.581 & 0.563 \\
\cline{2-10}
 & \multirow{3}{*}{ReCaLL} & w/o RCM & 0.777 & 0.776 & 0.702 & 0.472 & 0.473 & 0.471 & 0.404 \\
 & & RCM-base & 0.753 & 0.753 & 0.714 & 0.627 & 0.639 & 0.633 & 0.594 \\
 & & RCM & 0.753 & 0.753 & 0.715 & 0.626 & 0.639 & 0.632 & 0.594 \\
\midrule
\multirow{6}{*}{StarCoder2-3B} & \multirow{3}{*}{PPL} & w/o RCM & 0.584 & 0.566 & 0.503 & 0.484 & 0.461 & 0.455 & 0.422 \\
 & & RCM-base & 0.619 & 0.619 & 0.591 & 0.594 & 0.594 & 0.593 & 0.567 \\
 & & RCM & 0.618 & 0.618 & 0.590 & 0.594 & 0.593 & 0.593 & 0.566 \\
\cline{2-10}
 & \multirow{3}{*}{ReCaLL} & w/o RCM & 0.800 & 0.750 & 0.751 & 0.656 & 0.661 & 0.653 & 0.612 \\
 & & RCM-base & 0.759 & 0.759 & 0.732 & 0.768 & 0.767 & 0.764 & 0.739 \\
 & & RCM & 0.758 & 0.758 & 0.731 & 0.766 & 0.766 & 0.763 & 0.738 \\
\hline
\end{tabular}
}%
\end{table*}

Corresponding to the Type-1 to Type-3 mutation strategies introduced in \ref{subsec:RQ}, we design two stages in the reversion module. In each stage, semantically equivalent reversion is applied to remove mutation artifacts and gradually recover a more canonical form of the code. 

\textbf{Stage~1: LLM-based reversion for Type-1/Type-2 mutations.}
We first apply an LLM-based reversion step to mitigate formatting and lexical perturbations. Specifically, we use \texttt{Qwen2.5-Coder-32B}~\cite{hui2024qwen25codertechnicalreport} to transform each mutated snippet into a standardized form that follows common coding conventions. The model is prompted to improve the readability of the input code according to widely adopted style conventions. This step has two goals. First, it standardizes the formatting style of the code, removing inconsistencies caused by irregular indentation or spacing introduced by Mutation-T1. For example, the reversion process normalizes indentation width, removes redundant blank lines, standardizes spacing around operators and delimiters, and aligns control-flow blocks (e.g., \texttt{if}, \texttt{for}, \texttt{while}) into a consistent layout.

Second, the reversion step mitigates lexical perturbations introduced by Mutation-T2 by mapping identifiers to more common naming styles. In particular, the LLM is instructed to replace uncommon or artificially generated identifiers with clearer and more conventional alternatives while preserving variable roles. For instance, short or obfuscated variable names may be expanded into descriptive names, inconsistent naming styles (e.g., mixed camelCase and snake\_case) are unified, and redundant prefixes or suffixes are removed. While this process can not recover the exact original identifier names, it aims to produces a stable and consistent formatting and naming style that is commonly observed in real-world code repositories.

\textbf{Stage~2: Rule-based statement-level reversion for Type-3 mutations.}
To further handle statement-level mutations, we design a rule-based reversion stage that reverses the semantics-preserving reversion used in Mutation-T3. Concretely, we detect transformed statement patterns and convert them back to simpler and commonly used equivalents. The current implementation includes several categories of reversion rules, including:

\begin{itemize}
\item \textbf{Redundant boolean operations.} Remove unnecessary boolean constructs such as \texttt{cond and True}, \texttt{cond or False}, or double negations like \texttt{not(not cond)}.
\item \textbf{Equivalent conditional rewrites.} Simplify condition expressions that are syntactically expanded but semantically identical, such as replacing \texttt{if (cond == True)} with \texttt{if cond}.
\item \textbf{Loop structure normalization.} Convert alternative loop forms into simpler equivalents when possible, such as replacing \texttt{while True} with break conditions by an explicit conditional loop when the pattern is detectable.
\item \textbf{Redundant assignment or temporary variables.} Collapse trivial intermediate assignments that are introduced during mutation but do not affect program semantics.
\end{itemize}

After sequentially applying these reverse rules, this stage can reduce the structural noise introduced by statement-level mutation while preserving program semantics.

\textbf{RCM-base ablation.} To distinguish the contribution of the Stage-2 rules from the preceding canonicalization, we construct RCM-base by retaining the Type-1/Type-2 reversion stage and omitting the Stage-2 rule-based reversion. RCM-base therefore does not use the predefined statement patterns designed for Mutation-T3.

By combining LLM-based lexical reversion with rule-based structural reversion, the reversion module can effectively reduce a large portion of mutation artifacts. As a result, PDD methods are less likely to misclassify mutated member code as non-member due to semantically equivalent perturbations introduced by mutation.

\begin{figure}[t]
\centering
\includegraphics[width=\linewidth]{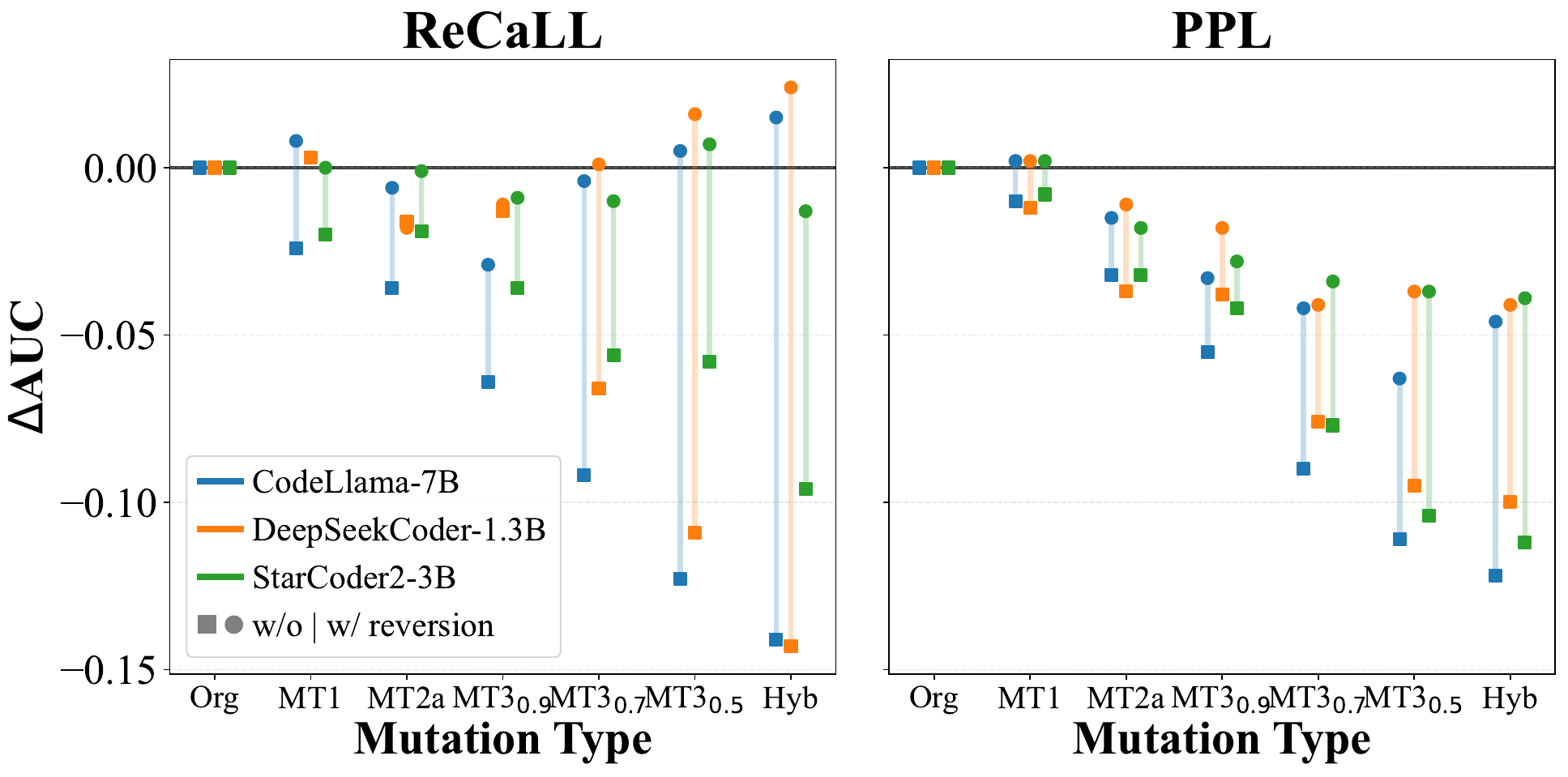}
\caption{The AUC degradation relative to the original \tool ($\Delta\text{AUC}$) on Setting 1, for ReCaLL (left) and PPL (right) across mutation types. For each mutation type, three representative models are reported under \textit{w/o reversion} and \textit{w/ reversion}. A smaller $\Delta\text{AUC}$ (in absolute value) indicates better recovery.}
\label{fig:rq5_reversion}
\end{figure}

\textbf{Evaluation setup.}
We evaluate the reversion module under \textbf{Setting 1}, where only the member samples are mutated, and the non-member set remains unchanged. For each mutation type, we compare two conditions: \textit{w/o reversion} (mutation only) and \textit{w/ reversion} (mutation followed by the reversion module). RCM and RCM-base are applied to both member and non-member inputs. We quantify the degradation of performance by
\begin{equation}
\Delta\text{AUC}=\text{AUC}_{\text{current}}-\text{AUC}_{\text{Org}},
\end{equation}
where $\text{AUC}_{\text{Org}}$ denotes the AUC on the original \tool, and $\text{AUC}_{\text{current}}$ denotes the AUC under the specific mutation type in either of the two conditions. A smaller $\Delta\text{AUC}$ (in absolute value) indicates less performance loss, meaning that the PDD method maintains stronger robustness under mutation.

\textbf{Computational cost.} Under the evaluation setup, the LLM-based canonicalization in Stage~1 requires an average of 0.782\,s per function, keeping the per-function inference time below one second. The rule-based reversion in Stage~2 requires an average of 16.3\,ms per function. These results show that the LLM-based stage has a controlled inference cost, while the subsequent rule-based stage is computationally efficient.

\textbf{Results.}
Figure~\ref{fig:rq5_reversion} retains the $\Delta\text{AUC}$ trends from the original Setting-1 evaluation. Table~\ref{tab:rq5_setting_1} complements the figure with an absolute-AUC ablation for PPL and ReCaLL across \texttt{CodeLlama-7B}, \texttt{DeepSeekCoder-1.3B}, and \texttt{StarCoder2-3B}.

The results show that RCM recovers a substantial part of the degradation caused by statement-level and hybrid mutations. For example, averaged across the three models, PPL increases from 0.492 to 0.599 under MT3$_{0.7}$ and from 0.449 to 0.573 under Hybrid after applying RCM-base. ReCaLL shows the same recovery pattern, increasing from 0.567 to 0.675 under MT3$_{0.7}$ and from 0.508 to 0.636 under Hybrid.
Across the eight evaluated CodeLLMs, RCM increases the Hybrid AUC by $\Delta\mathrm{AUC}=0.138$ for PPL (95\% CI $[0.135,0.142]$) and by $\Delta\mathrm{AUC}=0.125$ for ReCaLL (95\% CI $[0.121,0.129]$).

The effect is detector- and condition-dependent on the original and less complex mutations. On the original inputs, the three-model average changes from 0.595 to 0.617 for PPL, while ReCaLL changes from 0.796 to 0.757. Under MT1 and MT2a, ReCaLL remains close to or below its unprocessed score, whereas its recovery becomes clear for the stronger statement-level mutations.
On original, unmutated inputs, applying RCM to ReCaLL changes the AUC by $\Delta\mathrm{AUC}=-0.047$ (95\% CI $[-0.051,-0.043]$), indicating a measurable performance decrease in this condition.

Across the table, RCM-base and the full RCM have nearly identical AUCs: their largest difference is 0.002. The observed recovery therefore remains when the predefined Stage-2 Type-3 rules are removed, indicating that the preceding canonicalization accounts for most of the recovery in this evaluation.

\textbf{RCM under external mutation.} The Stage-2 rules are related to the statement patterns used by our Mutation-T3 module, so the original evaluation can partially reflect the reversibility of a known operator set. We therefore additionally evaluate RCM after an independently implemented transformation. Fully external Type-3 statement rewrites are difficult to construct with standard code transformation tools, which mainly provide lexical and formatting transformations. We use \texttt{python-minifier} 3.2.0~\cite{python_minifier} to compact the lexical surface and rename local identifiers, with AST transformations such as constant folding, assertion removal, and import combination. The same transformation and RCM preprocessing are applied to both member and non-member inputs in the fixed 300+300 subset.

The external transformation reduces CodeLlama-7B PPL and ReCaLL from 0.636 and 0.810 on the original inputs to 0.615 and 0.738, respectively; after RCM, they recover to 0.634 and 0.796. For StarCoder2-3B, ReCaLL changes from 0.799 to 0.741 under the external transformation and returns to 0.779 after RCM. PPL remains close to its original score (0.582, 0.600, and 0.595). These results show that RCM remains useful for an external transformation outside its predefined rule inventory, while the magnitude of recovery depends on the detector and model.

\begin{tcolorbox}[size=title]
{\textbf{Takeaway: RCM can recover performance under the evaluated code mutations. The RCM-base ablation shows that most of this benefit comes from Stage 1 canonicalization rather than the predefined Stage-2 reverse rules, and the external-transformation evaluation demonstrates that the benefit also extends beyond the original mutation operators.}}
\end{tcolorbox}

\section{Related Work}

{
\subsection{Training Data Detection for Language and Code Models}
Membership inference asks whether a record was used to train a model~\cite{shokri2017membershipinferenceattacksmachine}. For LLM pretraining, black-box detectors commonly score the target model's token likelihoods: perplexity and calibrated likelihoods~\cite{yeom2018privacyriskmachinelearning,274574,zhang2025pretrainingdatadetectionlarge}, low-probability tokens~\cite{shi2024detectingpretrainingdatalarge,zhang2025minkimprovedbaselinedetecting}, neighborhood comparisons~\cite{mattern2023membershipinferenceattackslanguage}, and relative conditional likelihoods~\cite{xie2025recallmembershipinferencerelative}. These methods avoid access to the pretraining corpus, but were developed mainly for natural-language data.

Code-specific studies use richer signals or stronger attacker assumptions. Gotcha combines the input, output, and ground-truth completion of a code model~\cite{yang2024gotchamodelusescode}; Buzzer extracts pretraining-task signals and calibrates sample difficulty~\cite{zhang2024buzzer}; CodeMI trains shadow models and an attack classifier for black-box completion models~\cite{wan2026codemi}; and AdvPrompt-MIA learns from code-specific adversarial prompt variations~\cite{jiang2025advprompt}. Syntax-aware scoring has also been explored for code attribution~\cite{li2026synprune}. Thus, existing methods cover complementary signals and access settings rather than a single standard PDD protocol.

Our work provides a systematic empirical study across these two lines of research. On a function-level, multi-language benchmark and multiple CodeLLMs, we evaluate transferable likelihood-based detectors together with CodeMI and AdvPrompt-MIA. We then extend the study beyond unchanged snippets by testing how code-clone transformations affect the operational membership definition and detector robustness.

\subsection{Code Clone Taxonomy}
Code-clone research commonly distinguishes four levels according to the edits needed to obtain one code fragment from another~\cite{roy2009comparison,10967509}. Type-1 clones differ only in formatting, Type-2 clones additionally change lexical elements such as identifiers, Type-3 clones contain statement-level additions, deletions, or substitutions, and Type-4 clones preserve semantics despite substantial syntactic differences. This taxonomy provides a natural vocabulary for discussing whether a transformed function should still be treated as related to an observed training example.

Applications of clone analysis to code models have examined whether generated completions reproduce training code~\cite{ciniselli2022cloning,alkaswan2024traces} and whether code is duplicated across training and evaluation datasets~\cite{lopez2024duplication}. Recent work further studies how semantically equivalent transformations affect membership inference for fine-tuned code models~\cite{yang2025semantically}. Our work makes clone type an explicit dimension of pretraining-data detection: the protocol operationalizes membership through Type-1 to Type-3 transformations and their combination, while Type-4 semantic equivalence remains outside the evaluated scope. This design complements existing work by comparing multiple PDD methods across progressively larger transformations and by evaluating RCM as preprocessing for this setting.
}

\section{Threats to Validity}

\noindent{\textbf{Limited Models.}} Although we have studied eight representative CodeLLMs across different model families and sizes, the landscape of code models is rapidly evolving. There may be other models—either more recent, proprietary, or domain-specific—that exhibit different behaviors and could potentially alter the conclusions drawn in this study. Future work should include a broader range of models to validate generalizability.

\noindent{\textbf{Limited Programming Languages.}} This study focuses on three widely used programming languages to maintain a controlled and comparable evaluation environment. However, programming languages differ significantly in syntax, structure, and idiomatic usage. As such, the effectiveness of training data detection methods may vary when applied to other languages, particularly low-resource or domain-specific ones. Expanding the language coverage would help assess the robustness and adaptability of detection methods.

\noindent{\textbf{Temporal Distribution.}} The member and non-member sets are constructed under different temporal conditions. Member samples are drawn from publicly released pretraining corpora, while non-member samples are collected from repositories and commits that postdate the evaluated models. This construction may introduce temporal novelty signals in addition to membership-related signals~\cite{duan2024do}. Our temporal analysis shows that detection performance changes across non-member collection periods.

\noindent{\textbf{Membership Labels.}} Membership labels are assumed rather than confirmed: the corpus sources provide high-confidence evidence for some model families, but the public training documentation does not establish that the CodeLlama models were trained on the member sources used here. Likewise, post-cutoff provenance makes non-member inclusion less likely but cannot rule out earlier copies or near duplicates. Signature-level deduplication removes exact function overlaps, but does not exclude Type-1/2/3 near duplicates across the two sets.

\noindent\noindent{\textbf{Limited Clone-Type Scope.}} Our mutation protocol evaluates formatting (Type-1), lexical (Type-2), and semantics-preserving statement-level (Type-3) transformations, together with their hybrid combination. It does not evaluate Type-4 semantic clones, whose syntax may differ substantially despite equivalent functionality. Therefore, our empirical membership definition and conclusions apply to the evaluated Type-1 to Type-3 transformations and should not be generalized to Type-4 semantic equivalence.

\noindent{\textbf{Limited Dataset.}} While the dataset used is relatively large-scale and designed to reflect diverse real-world coding scenarios, it is not exhaustive. Certain edge cases, rare coding patterns, or specific mutation strategies may not be adequately captured. This incompleteness may influence the generality of the conclusions, especially when evaluating models trained on significantly different or proprietary corpora. Future studies could benefit from more comprehensive, community-curated datasets covering a broader spectrum of code variations.

\section{Conclusion}
We curate a multi-language code benchmark \tool for training data detection and conduct a systematic evaluation of nine state-of-the-art training data detection methods on eight popular CodeLLMs. The study reveals varying effectiveness in identifying code membership.
The results show that ReCaLL achieves the best performance across all settings, including different models, mutation strategies, code lengths, and programming languages. However, its performance remains suboptimal, especially in mutation scenarios and with certain programming languages.
We also design RCM to enhance the performance of existing PDD methods.
Through this study, we aim to encourage more research into the training data detection task for code data, which will help establish a foundation for future advancements in the compliant deployment of CodeLLMs.


\bibliographystyle{IEEEtran}
\bibliography{sample-base}

\end{document}